\documentclass[aps,floatfix]{revtex4}
\usepackage{amssymb,hyperref,hypcap,
ae,tikz,color}
\usepackage{latexsym,textcomp,slashed,epsfig}
\usepackage[T1]{fontenc}
\usepackage{verbatim}

\usepackage{dcolumn}
\usepackage{bm}
\usepackage{amsmath}
\usepackage{graphicx}
\usepackage{listings}
\usepackage[T1]{fontenc}
\usepackage{xcolor}
\usepackage{lmodern}
\usepackage{listings}
\usepackage[percent]{overpic}

\hypersetup{
	colorlinks=true,
	citecolor=blue,
    linkcolor=magenta,
    urlcolor=blue
	}

\begin{document}

\title{{\bf On the propagation of gravitational waves in matter-filled Bianchi I universe}}

\author{\bf{Sucheta Datta, Sarbari Guha and Samarjit Chakraborty}}

\affiliation{\bf Department of Physics, St.Xavier's College (Autonomous), Kolkata 700016, India}

\maketitle

\section*{Abstract}
In this paper we apply the Regge-Wheeler formalism to study the propagation of axial and polar gravitational waves in matter-filled Bianchi I universe. Assuming that the expansion scalar $ \Theta $, of the background space-time, is proportional to the shear scalar $ \sigma $, we
solved the background field equations in the presence of matter (found to behave like a stiff fluid). We then derive the linearised perturbation equations for both the axial and polar modes. The analytical solutions in vacuum spacetime could be determined in an earlier paper \cite{GD1} in a relatively straightforward manner. However, here we find that in the presence of matter, they require more assumptions for their solution, and bear more involved forms.
As compared to the axial modes, the polar perturbation equations contain far more complicated couplings among the perturbing terms. Thus we have to apply suitable assumptions to derive the analytical solutions for some of the cases of polar perturbations. In both the axial and polar cases, the radial and temporal solutions for the perturbations separate out as products. We find that the axial waves are damped owing to the  background anisotropy, and can deform only the azimuthal velocity of the fluid. In contrast, the polar waves must trigger perturbations in the energy density, the pressure as well as in the non-azimuthal components of the fluid velocity. Similar behaviour is exhibited by axial and polar gravitational waves propagating in the Kantowski-Sachs universe \cite{GD2}. Our work is in contrast to the work done in \cite{SYK}, where the authors analysed anisotropic universes modelled by Kasner spacetime and Rindler wedges using the method of gauge-invariant perturbations in the RW gauge.

\bigskip

KEYWORDS: Regge-Wheeler gauge, Bianchi I spacetime, Axial Gravitational waves, Polar Gravitational waves

\section{Introduction}
The detection of gravitational waves (GWs) has confirmed the last prediction of Einstein's General Theory of Relativity.
The geometry of a spacetime is related to its matter content through Einstein's equations and GWs are obtained as their solutions under perturbations in the linearised approximation.

The present-day isotropic and homogeneous universe is described by the FLRW model. Although this model offers suitable explanation for the current state of the universe, certain observations indicate the existence of cosmological anisotropies. So anisotropic models and the evolution of perturbations in these backgrounds have gained importance. Of  the various anisotropic but homogeneous models \cite{ELLIS}, the Bianchi I (B-I) universe \cite{BIANCHI} is the simplest one. The B-I model has different scale factors along different directions. In such a universe filled with matter obeying an equation-of-state (EoS) $ p = \gamma \rho $, where the EoS parameter $ \gamma < 1 $, any initial anisotropy quickly dies away and eventually evolves into a FLRW universe \cite{JACOBS}. The B-I spacetime is therefore a potential candidate for modelling the early universe.

Gravitational perturbations in homogeneous anisotropic universes (including B-I universes) have been extensively studied. Such investigations are of interest to understand how anisotropy influences the gravitational perturbations. Detection of primordial GWs can shed light on the nature of the early universe. Hu \cite{HU} pointed out that perturbations in such universes can be treated as a first approximation to exact inhomogeneous anisotropic solutions in the chaotic cosmology, and the back-reaction of GWs 
may considerably change the dynamics of the early universe. 
He showed the decoupling of the two linear polarizations of the GWs. In contrast to the FLRW universe, where the two polarizations are decoupled, each being equivalent to a minimally coupled massless scalar field \cite{CHOref}, these become coupled in an expanding anisotropic universe \cite{CHO}. The paper \cite{MIEDEMA93} dealt with general perturbations of the B-I spacetime. Others \cite{ADAMS1} introduced a non-perturbative exact formalism for GWs travelling through Bianchi I-VII spacetimes. The property obtained in \cite{ADAMS1} was generalized in \cite{ADAMS2} that there exist solutions in which the inhomogeneity initially dominates the structure of the cosmic singularity but later evolves into GWs propagating over the more homogeneous background. 


Among the perturbation formalisms developed to study GWs, the Regge-Wheeler perturbation scheme is a relatively simple one. In their investigations on Schwarzschild black hole perturbations, Regge and Wheeler \cite{RW} decomposed the perturbing elements in terms of spherical harmonics, and incorporated the Regge-Wheeler (RW) gauge in order to extract a single Schr\" odinger-type differential equation for the axial as well as polar perturbations. The solutions come out in the form of products of four factors, each being a function of only one coordinate ($ t $, $ r $, $ \theta $, $ \phi $). Subsequently, this procedure has been used in various articles \cite{ZER}-\cite{VIA}. Zerilli made corrections to the polar equation of \cite{RW} in \cite{ZER} and studied the problem of a particle falling into a Schwarzschild black hole \cite{ZERprd70}. Vishveshwara \cite{VISH} used the RW gauge in Kruskal coordinates to examine the stability of the Schwarzschild exterior metric.
Metric perturbations have also been discussed in \cite{CHANDRA}. In Ref.\cite{QI}, the solution to the Schwarzschild perturbations was found to describe an outgoing axial wave corresponding to a special case of Chandrasekhar's perturbations. Anderson \textit{et al.} derived an asymptotic gauge \cite{ANDER} to transform the metric from RW gauge to Chandrasekhar-Esposito gauge \cite{CHANDRA2}. The exact axial solutions in the Schwarzschild background as calculated by Fiziev \cite{FIZ} are given by the confluent Heun's functions. Much simpler solutions for RW equation have been reported later \cite{HASSAN}. In the de Sitter spacetime \cite{BINI}, the set of perturbation equations, on using the RW formalism, reduces to a single second-order differential equation of the Heun-type for both electric and magnetic multipoles.

The RW procedure has been applied to the axial gravitational perturbations in the FLRW model by Malec and Wylezek \cite{M1}. 
Kulczycki and Malec \cite{M2} investigated the polar GWs which are found to cause perturbations in the density and non-azimuthal components of the velocity of the material medium, and are hence responsible for the evolution of matter inhomogeneities and anisotropies. On the other hand, since the initial data can be so adjusted as to decouple from matter, axial waves can perturb only the azimuthal velocity and trigger local cosmological rotation.
In such spacetimes, if the initial profiles are not smooth, then the axial wave pulses bring about rotation of the radiation fluid, leading to the memory effect \cite{M3}.
The propagation of axial and polar waves in FLRW universes using the RW formalism has been studied by Sharif and Siddiqa in the $ f(R,T) $ gravity \cite{SH1, SH2} and by Salti \textit{et al.} in Rastall gravity \cite{SALTI1, SALTI2}. In the Starobinsky model of $ f(R) $ gravity, Ref. \cite{SIDD} derives analytical solutions to the axial perturbations in the radiation era and the de Sitter stage. Rostworowski \cite{ROST1} has suggested using the RW formalism to study perturbations, as an alternative to the standard Bardeen's formalism of gauge-invariance \cite{BARDEEN}.
A recent work \cite{NAYAK} determines the equivalent of the RW equation, for generalized McVittie metric. 
Ref.\cite{LENZI} examines all possible (gauge-invariant) master functions and equations for the perturbations to vacuum spherically-symmetric spacetimes. 
Axial GWs in flat FRW universe and in the $ f(R, T^{\phi}) $ gravity have been presented in \cite{SIDD2}.

Axial and polar perturbations have been investigated in the gauge-invariant framework in \cite{GMG, MGG, GS, CL1}. With a $2+2$ decomposition of the metric, Gundlach and Martin-Garcia \cite{GMG} analysed non-spherical perturbations of a spherically symmetric, time-dependent spacetime to study the generation of GWs. They have found three axial and seven polar gauge-invariant matter perturbations. 
Martel and Poisson \cite{MAR} presented the Schwarzschild metric perturbations in the covariant, gauge-invariant formalism. The metric perturbation theory was analysed by Chandrasekhar \cite{CHAND} also.

Clarkson \textit{et al.} \cite{CL1} have produced a complete set of master equations for the LTB dust model. The decomposition of any perturbation into scalar, vector and tensor (SVT) modes, all evolving independently, which is feasible in the FLRW model, cannot be done in the LTB model, where the modes get coupled. However, they can be decoupled into two independent modes - axial (or odd) and polar (or even), classified according to the nature of their transformation on spherically symmetric surfaces.
As non-trivial symmetric, transverse and trace-free rank-2 tensors cannot exist on $ S^2 $, further decomposition into tensor modes is not possible here. The master equation of the polar waves is numerically solved in \cite{CL2}.
Using the 1+1+2 covariant decomposition of spacetime, Keresztes \textit{et al.} \cite{BR1} have carried out a study of the perfect-fluid perturbations of the Kantowski-Sachs universe with vanishing vorticity and a positive cosmological constant and generalised the analysis for LRS class-II cosmologies in \cite{BR2}.
Although there arise four propagation equations for gravitational perturbations, it has been clarified in \cite{BR1} that GWs possess two degrees of freedom, the `+' and `$ \times $' polarisations.
However, in modified theories of gravity, there can be upto six polarisations of GWs, the additional degrees of freedom occurring due to scalar modes, whose presence or absence are observer-dependent \cite{ALVES1}. These polarisations have been studied extensively in \cite{CORDA1}-\cite{GONG3}.


Several studies on axial and polar GWs employing the RW perturbation scheme have been performed with FLRW metric as the background \cite{M1, M2, M3, SH1, SH2, ROST1, ROST2, SALTI1, SALTI2}. But the same has not been widely discussed for an anisotropic spacetimes. In the article \cite{SYK}, the authors explored GWs in anisoptropic universe using the method of gauge-invariant perturbations in the RW gauge \cite{GS} to decompose the metric tensor of the Kasner spacetimes (and Rindler wedges), possessing two-dimensional plane symmetry. They expanded the Einstein-Hilbert Lagrangian to second order in metric perturbations to obtain the solutions. Aiming to focus on the quantum aspects of GWs, they introduced two decoupled master variables which represent the gravitational degrees of freedom, and each of which is equivalent to the action of a massless scalar field in the corresponding background. The odd-parity and even-parity perturbations were obtained by solving the massless Klein-Gordon equation in each spacetime considered by them. Recently, we studied the propagation of GWs in Kantowski-Sachs background \cite{GD2}. Our previous work on Bianchi I (B-I) background also invoked RW gauge \cite{GD1}. In these two articles, we determined the solutions of the vacuum perturbation equations analytically.

In the present paper, we aim to find complete analytical solutions when the B-I spacetime is no longer vacuous, but filled with matter, which is found to obey the stiff fluid EoS. Here the system of perturbation equations become more involved and require special conditions to be solved. In contrast to \cite{SYK}, our work is based on the conventional approach followed by Malec and others. The axial and polar perturbations are decomposed in terms of spherical harmonics and simplified using the RW gauge. We have been able to determine solutions to Einstein equations in Bianchi I universe in the presence of such perturbations, both in matter- filled and vacuum cases. The explicit dependence of each perturbing element on $t$ and $r$ has been shown graphically.

Our paper is organised as follows: Sec.II presents the B-I background metric, the corresponding Einstein's equations and solutions in terms of scale factors in the presence or absence of matter. In Sec.III, the Regge-Wheeler gauge is introduced. Sec.IV carries a note on the perturbed energy-momentum tensor and four-velocity of the fluid. In Sec.V, we concentrate on the axial GWs. The linearised field equations for the axially-perturbed background are derived in the presence of matter and then solved analytically under certain assumptions. The vacuum case is touched upon in brief.  Sec.VI deals with the polar modes. The perturbation equations in the presence of matter (specifically stiff fluid), and in vacuum, followed by their analytical solutions in particular cases are obtained. We conclude with an analysis of our results and some remarks in Sec.VII. Throughout this paper, we will use geometrized units, i.e., $ 8\pi G = c = 1 $, and overdots and primes to represent derivatives w.r.t. $ t $ and $ r $ respectively. 

\section{The unperturbed background metric and field equations}

Exact solutions for homogeneous spacetimes in GR belongs to either Bianchi types or the Kantowski-Sachs (KS) model \cite{AKRC,KS}. Among the LRS cosmologies of class II, the KS spacetime has positive curvature (2D scalar curvature), while those with zero and negative curvature are respectively the Bianchi I/ VII$ _0 $ (including flat Friedmann universes), and the Bianchi III models \cite{BR1, KATORE}. The general form of the anisotropic line element can be written as \cite{SHAMIR1006, BR1} :
\begin{equation}
ds^2 = dt^2 -a^2(t)dr^2 -b^2(t)[d\theta ^2 + f_{\mathcal{K}}(\theta) d\phi ^2],
\end{equation}
$ \mathcal{K} $ being the spatial curvature index of the spacetime.\\
For $ \mathcal{K} =0 $, $ f_{\mathcal{K}}(\theta) =\theta ^2 $, the universe is classified as Bianchi I.\\
For $ \mathcal{K} =1 $, $ f_{\mathcal{K}}(\theta) =\sin ^2 \theta $, the model is closed, and named after Kantowski and Sachs.\\
For $ \mathcal{K} =-1 $, $ f_{\mathcal{K}}(\theta) =\sinh ^2 \theta $, the model represents the semi-closed Bianchi III space-time.\\

This paper is devoted to the study of the propagation of GWs in matter-filled Bianchi I spacetime (i.e., for $ \mathcal{K} =0 $), although, for polar perturbations we will also examine the propagation in absence of matter. To begin with, the corresponding line element is defined in spherical polar coordinates by:
\begin{equation} \label{1}
ds^2= dt^2- a^2(t) dr^2 -b^2(t) d\theta ^2 -b^2 (t) \theta ^2 d\phi ^2 ,
\end{equation}
where $ a(t) $ and $ b(t) $ are the scale factors representing the expansion along the directions parallel and the perpendicular to the radial direction respectively. Considering the spacetime to be filled with a perfect fluid having four-velocity $u^\alpha$, energy density $ \rho $ and pressure $ p $, the energy-momentum tensor is given by:
\begin{equation} \label{1a}
T_{\mu\nu}= (\rho+p) u_{\mu}u_{\nu} -p g_{\mu\nu}.
\end{equation}
The background field equations are obtained as:
\begin{equation}\label{2}
\frac{2 \dot{a}\dot{b}}{ab} + \frac{\dot{b}^2}{b^2} =\rho_0, \qquad
-\left( \frac{2\ddot{b}}{b} +\frac{{\dot{b}}^2}{b^2} \right)= p_0, \qquad  \textrm{and} \qquad
-\left( \frac{\ddot{a}}{a} +\frac{\ddot{b}}{b} +\frac{\dot{a} \dot{b}}{ab} \right)= p_0.
\end{equation}
The continuity equation is represented by:
$ \dot{\rho_0} + \left( \dfrac{\dot{a}}{a} + \dfrac{2\dot{b}}{b}\right) (\rho_0 +p_0) =0. $
The subscript `0' refers to the physical quantities (here, the energy density and pressure of the fluid) associated with the background metric \eqref{1}.

From the above we find that the three equations in \eqref{2} along with the continuity equation are coupled in such a way that they cannot be solved unless we assume an additional condition. We therefore proceed to find a relation between the two scale factors by assuming the physical condition that the background spacetime remains anisotropic throughout its entire evolution.

\subsection{Relation between the two scale factors in Bianchi I spacetime}
For the above metric \eqref{1}, we find that the volume expansion $ \Theta $ and shear scalar $ \sigma $ are given by
\begin{equation}\label{E1}
\Theta = \frac{\dot{a}}{a} + \frac{2\dot{b}}{b},  \hspace{0.5cm} \text{and} \hspace{0.5cm}  \sigma ^2 = \frac{1}{3}\left( \frac{\dot{a}}{a} - \frac{\dot{b}}{b} \right)^2 .
\end{equation}
It is known that a cosmological model remains anisotropic throughout its evolution if the ratio of the shear $ \sigma $ to the expansion $ \Theta $ remains constant \cite{GRON2, ROY1, ROY2, BAG, BALI}. Thus the expansion scalar must be proportional to the shear scalar, and we can assume that
\begin{equation}\label{E2}
a=b^n,
\end{equation}
where $ n $ is an arbitrary real number and $ n \neq 0, 1 $ for non-trivial solutions. 
Substituting this in Eqn.\eqref{E1} gives
\begin{equation}
\frac{\sigma ^2}{\Theta ^2} =\frac{1}{3}\left( \frac{n-1}{n+2} \right) ^2.
\end{equation}
Clearly, this ratio is constant for any value of $ n $. 

\subsection{Solutions for the background equations in presence of matter}
In view of the relation \eqref{E2}, the background equations \eqref{2} are solved by
\begin{equation}\label{3c}
b(t)= [(n+2)(k_1 t + k_2)]^{\frac{1}{n+2}},
\end{equation}
where $ k_1 $ and $ k_2 $ are the integration constants.  This is equivalent to the corresponding expression determined by Shamir \textit{et al.} in the $ f(R,T) $ theory \cite{SHAMIR1507}. Assuming $ b(t)=0 $ at $ t=0 $, $ k_2 $ vanishes and we are left with
\begin{equation}\label{3d}
 b(t)= [(n+2)k_1 t]^{\frac{1}{n+2}},  \hspace{0.5cm} \text{and hence} \hspace{0.5cm} a(t)= [(n+2)k_1 t]^{\frac{n}{n+2}}.
\end{equation}
Substituting these expressions for $ a(t) $ and $ b(t) $ in equations \eqref{2}, we find that
\begin{equation}\label{3e}
\rho_0(t) = p_0(t)= \dfrac{2n+1}{[(n+2)t]^2},
\end{equation}
which represents a stiff fluid \cite{SHAMIR1507}. This expression also indicates that $ n=-1/2 $ for vacuum solutions where $ \rho_0(t) = p_0(t)= 0 $.

We know that the equation of state of a stiff perfect fluid is given by :
\begin{equation}\label{sf}
p_0 = \rho_0.
\end{equation}
This is the extreme relativistic limit for a perfect fluid, when the speed of sound becomes equal to that of light. This was first suggested by Zel'dovich \cite{ZELD}, who considered the early universe to be composed of a cold gas of baryons behaving like a stiff fluid. Its governing equations have the same characteristics as that of the gravitational field \cite{WESSON}.
After the cosmic explosion, when the universe was characterised by high densities, the matter content could have been stiff \cite{ZELD, BARROW, CHAVANIS}.
Various aspects of Bianchi I spacetimes containing stiff fluid have been mentioned in \cite{BALI2}, \cite{BANERJEE}. Since the stiff fluid in the FLRW universe exhibits faster decrease in density than radiation and matter, it plays a significant role in the early phase. This helped Dutta and Scherrer \cite{DS} to numerically compute the effect of the stiff fluid density on the primordial abundances of light elements.
Although GWs propagating through dust or through fluids with a realistic equation of state are somewhat problematic to examine, exact radiative solutions can be found if the medium is considered as a stiff fluid \cite{GRIF1}-\cite{AG}. In \cite{MATHEW2}, the effective equation-of-state of the Zel'dovich fluid is shown to start evolving from stiff nature, pass through a pressure-less state and eventually tend towards de Sitter epoch. The fluid when combined with decaying vacuum energy \cite{MATHEW3} yields the age of the universe in agreement with the observations.


According to \cite{WAINBOOK}, a Bianchi I model defined by
\begin{equation}\label{4}
ds^2 = -dt^2 + t^{2p_1}dx^2 + t^{2p_2}dy^2 + t^{2p_3}dz^2 ,
\end{equation}
and characterised by a non-tilted perfect fluid with an equation of state: $ p_0= (\gamma-1)\rho_0 $, $ \gamma $ = constant, has Jacobs' stiff perfect fluid solutions if the following relations hold:
\begin{equation}\label{4a}
\begin{split}
p_1 + p_2 + p_3 =1, \hspace{0.5cm} p_1^2 + p_2^2 + p_3^2 <1,  \hspace{0.5cm}  
\rho_0= \frac{1}{2}(1 -p_1^2 - p_2^2 - p_3^2) t^{-2}, \hspace{0.5cm} \gamma =2.
\end{split}
\end{equation}
In view of this, in order to have the matter content corresponding to the line element \eqref{1} to be described by the equation of state of a stiff fluid, we propose that
\begin{equation}\label{4b}
a(t) \propto t^{p_1}, \hspace{0.5cm} b(t) \propto t^{p_2}, \hspace{0.5cm} p_2=p_3.
\end{equation}
The first two conditions in \eqref{4a} will be satisfied if $ p_1=1/2 $, and $ p_2=1/4 $ in our case. In that case, we have
\begin{equation}\label{4e}
a \propto t^{1/2} \hspace{0.5cm} \text{and} \hspace{0.5cm} b \propto t^{1/4}.
\end{equation}
This means that, according to \eqref{E2} we must have $n=2$. Using these values of $ p_1 $ and $ p_2 $ along with the other two relations in \eqref{4a}, we find that
\begin{equation}\label{4c}
\rho_0 =p_0 =\dfrac{1-\frac{3}{8}}{2t^2} =\dfrac{5}{16t^2}.
\end{equation}
Equating equations \eqref{3e} and \eqref{4c} gives  $ n=2, \hspace{0.2cm} \text{or}, \hspace{0.2cm} n=2/5. $
Considering this result and the value of $ n $ obtained in \eqref{4e}, we can say that the solution $ n=2 $ is appropriate. Hence, from Eqn.\eqref{3d}, we have
\begin{equation}\label{4f}
\hspace{0.5cm}
b(t)= (4k_1 t)^{\frac{1}{4}} = \kappa t^{\frac{1}{4}}, \hspace{0.5cm} a(t)= b(t)^2 = \kappa ^2t^{\frac{1}{2}},
\hspace{1cm} \kappa = (4k_1)^{\frac{1}{4}} =  \text{constant}.
\end{equation}

\subsection{Solutions for the background equations in vacuum}

Solving the equations in the set \eqref{2} for the vacuum case, using the relation \eqref{E2}, we obtain
\begin{equation}\label{6}
b(t)= K t^{2/3}, \quad \text{and} \quad a=b^{-1/2}.
\end{equation}
Here $ K $ is the integration constant for vacuum solutions. These scale factors are used for our calculations in the case of polar perturbations, as the case of axial perturbations in vacuum has already been studied in details in our previous work \cite{GD1}.

\section{The perturbed metric in the Regge-Wheeler gauge}

Gravitational waves propagating in the Bianchi I background are represented by small perturbations $ h_{\mu\nu} $, and the perturbed metric is defined as:
\begin{equation}\label{7}
g_{\mu\nu} = g_{\mu\nu}^{(0)} + e h_{\mu\nu} + \mathcal{O}(e^2).
\end{equation}
Here, $ g_{\mu\nu}^{(0)} $ is the background metric \eqref{1}. The smallness of the magnitude of the perturbations is indicated by the parameter $ e $, and all terms of $ \mathcal{O}(e^2) $ are to be neglected in the calculations.

As explained in Ref.\cite{RW}, the components of the perturbation matrix transform differently under a rotation of the frame about the origin. For example, $ h_{00} $, $ h_{01} $, $ h_{11} $ transform like scalars, $ (h_{02} $, $ h_{03} $) and ($ h_{12} $, $ h_{13} $) transform like vectors, and $ h_{22} $, $ h_{23} $ and $ h_{33} $ like a second-order tensor. These are decomposed in terms of spherical harmonics $ Y_{lm} $ ($ l $ is the angular momentum and $ m $ is its projection on the $ z $-axis), and grouped according to odd or even parity. It is then found that the axial waves are given by three unknown functions of $ r $, and the polar waves by seven unknown functions. At this point, the Regge-Wheeler gauge is introduced to find the canonical form of the axial and polar waves. Subsequently, the $ t $ and $ r $-solutions separate out as product in the final expressions.

For odd (or axial) waves, the matrix $ h_{\mu\nu} $ has only two non-zero components \cite{M2} represented by: \begin{equation}\label{7a}
h_{t \phi} =h_0(t,r) \sin \theta (\partial_{\theta}Y) \hspace{0.4cm} \text {and} \hspace{0.4cm}
h_{r \phi} =h_1(t,r) \sin \theta (\partial_{\theta}Y).
\end{equation}
Thus the background metric \eqref{1} in the presence of the axial perturbations is given by:
\begin{equation}\label{7b}
ds^2= dt^2- a^2(t) dr^2 -b^2(t) d\theta ^2 -b^2(t) \theta ^2 d\phi ^2
+2eh_0(t,r) \sin \theta (\partial_{\theta}Y) dtd\phi +2eh_1(t,r) \sin \theta (\partial_{\theta}Y) drd\phi + \mathcal{O}(e^2).
\end{equation}
Here, the spherical harmonics $ Y_{lm}(\theta, \phi) $ are denoted by $ Y $, with $m=0$. For all values of $ m $, the radial equation remains the same. Hence, $ m $ is chosen to be zero and the $ \phi $-dependence of $ Y $ is removed \cite{RW}. For wavelike solutions, $ l \geq 2 $ \cite{SH1}. Further, the spherical harmonics follow the relation:
\begin{equation}\label{sph}
\partial_{\theta} \partial_{\theta} Y = -l(l+1)Y - \cot\theta (\partial_{\theta} Y).
\end{equation}

Coming to the even (or polar) waves, it is found that there are a number of non-zero components of $ h_{\mu\nu} $ \cite{RW}. Adopting the Gerlach-Sengupta \cite{GS} formalism, which has later been developed by Gundlach and Martin-Garcia \cite{GMG}, Clarkson and others \cite{CL1} have proposed the general form of the polar perturbations as:
\begin{equation}\label{7e}
h_{\mu\nu} =
\begin{pmatrix}
 (\chi+\psi -2\eta)Y  &\zeta Y  &0  &0 \\
 \zeta Y  &(\chi+\psi)Y  &0  &0 \\
 0  &0  &\psi Y  &0 \\
 0  &0  &0  &\psi Y
\end{pmatrix}.
\end{equation}
$ \chi $, $ \psi $, $ \zeta $ and $ \eta $ are all functions of $ t $ and $ r $, and equivalent to the gauge-invariant variables introduced in \cite{GMG,MGG,GS}. The matrices in \eqref{7a} and \eqref{7e} correspond to magnetic and electric multipoles respectively \cite{BINI}. The polar perturbations for the Bianchi I background \eqref{1} are therefore given by:
\begin{equation}\label{7f}
\begin{split}
ds^2= [1+ e \left\lbrace \chi(t,r)+\psi(t,r) -2\eta(t,r) \right\rbrace Y] dt^2 + 2e\zeta(t,r)Y dtdr 
+[-a^2(t) + e \left\lbrace \chi(t,r)+\psi(t,r) \right\rbrace Y] dr^2 \\
+[-b^2(t) +e\psi(t,r)Y] d\theta ^2
+[-b^2(t) +e \psi(t,r)Y] \theta ^2 d\phi ^2  + \mathcal{O}(e^2).
\end{split}
\end{equation}
The value of $ \eta $ in the above equations may or may not be zero. $ \eta (t,r) $ is non-zero in the $ (0-0) $ element of the corresponding perturbation matrix for the LTB background \cite{CL1}. The constraint $ \eta =0 $ does not hold for the field equations in \cite{CL1} while considering large-angle fluctuations, where $ l=0 \: \textrm{or} \: 1$. But $ \eta $ must vanish in the FLRW background \cite{M2, ROST2}. Also, in \cite{GMG, MGG}, for $l\geqslant 2$, one has $\eta =0$. However, gauge-invariance is no longer valid \cite{GMG} for the polar $l=0, \: \textrm{and} \: 1$ cases. For $l=1$, there exist no dipole tensorial spherical harmonics, and hence $ \eta $ is no longer zero. Moreover, due to the missing tensorial components, the gauge-invariant variables defined for $l\geqslant 2$ become partially gauge-invariant, leaving one degree of freedom to be fixed \cite{CL1, MEYER}. For $l=0$, there are two degrees of freedom (see \cite{MGG}, Appendix A).  Therefore, additional constraints are required for gauge fixing.


\section{The perturbed energy-momentum tensor}

Let us consider the perturbations in the energy density and pressure of the fluid. These terms can be written (adopting the method in the FRLW case \cite{M2, SH1, SH2}) in the following way:
\begin{eqnarray}
\rho = \rho_0 (1+ e\cdot \Delta(t,r) Y) +\mathcal{O}(e^2),   \hspace{1cm} 
p = p_0 (1+ e\cdot \Pi(t,r) Y) +\mathcal{O}(e^2), \label{8b}
\end{eqnarray}
where $ \Delta(t,r) $ and $ \Pi(t,r) $ are the perturbations in the energy density and pressure, respectively. Since the equation of state relates the background energy density $ \rho_0 $ and pressure $ p_0 $, hence $ \Delta(t,r) $ and $ \Pi(t,r) $ are also related to each other. Moreover, the four-velocity of the fluid has to be taken into account while incorporating the perturbations. The fluid may or may not be co-moving with the unperturbed cosmological expansion of the universe.
The perturbed components of the fluid four-velocity $ u_\alpha =(u_0,u_1,u_2,u_3) $ are defined as in \cite{M2}:
\begin{eqnarray}
u_0 = \frac{2 g_{00}^{(0)} +e h_{00}}{2} +\mathcal{O}(e^2), \hspace{1.5cm} 
u_1 = e a(t) w(t,r) Y +\mathcal{O}(e^2), \\  
u_2 = e v(t,r) (\partial_{\theta}Y) +\mathcal{O}(e^2), \hspace{1cm} 
u_3 = e U(t,r) \sin \theta (\partial_{\theta}Y) +\mathcal{O}(e^2), \label{8f}
\end{eqnarray}
such that
\begin{equation}\label{N}
u_{\mu}u^{\mu} = 1+ \mathcal{O}(e^2).
\end{equation}
Here, $ h_{00} $ is the $ (0-0) $ element of the perturbation matrix.
Substituting the respective expressions in Eqn.\eqref{1a}, one can determine the non-zero components of the perturbed energy-momentum tensor. 




\section{Axial perturbations: Equations and Solutions}

First we derive the axial perturbation equations for the matter-filled Bianchi I background using the RW gauge. Subsequently, we look for their solutions to express the perturbing terms as products of functions of $ t $ and $ r $.

\subsection{Perturbation equations in presence of matter: }

The field equations for the axially perturbed metric \eqref{7b} for a perfect fluid of energy density $ \rho_0 $ and pressure $ p_0 $, are:
\begin{eqnarray}
\frac{2 \dot{a}\dot{b}}{ab} + \frac{\dot{b}^2}{b^2} =\rho_0 (1+e\Delta Y),  \label{11a} \hspace{6.0cm}  \\
-\left( \frac{2\ddot{b}}{b} +\frac{{\dot{b}}^2}{b^2} \right) = p_0(1+e\Psi Y),  \label{11b} \hspace{5.8cm}  \\
-\left( \frac{\ddot{a}}{a} +\frac{\ddot{b}}{b} +\frac{\dot{a} \dot{b}}{ab} \right) = p_0(1+e\Psi Y), \label{11c} \hspace{5.5cm}  \\
(\rho_0 + p_0) e a w Y =0,  \label{11e} \hspace{6.5cm}  \\
(\rho_0 + p_0) e v (\partial_{\theta}Y) =0,  \label{11f} \hspace{6.4cm}  \\
\frac{e}{2} \sin\theta (\partial_{\theta}Y) \left[
\frac{h_0''}{a^2} -\frac{\dot{h}_1'}{a^2}  +\frac{2 \dot{b}h_1'}{a^2 b} +\frac{2 \ddot{a}h_0}{a} +\frac{2 \ddot{b}h_0}{b}
+\frac{2 \dot{a} \dot{b} h_0}{ab}  -\frac{h_0}{b^2 } \left\lbrace  l(l+1) \right\rbrace  \right] =\left[ (\rho_0 +p_0)U -p_0 h_0 \right] e \sin \theta (\partial_{\theta}Y),  \label{11g} \\
-\frac{e}{2} \sin\theta (\partial_{\theta}Y) \left[ \ddot{h}_1 -\dot{h}_0' -\frac{\dot{a}\dot{h}_1}{a}  +\frac{\dot{a}h_0'}{a} -\frac{2 \dot{b} h_0'}{b} -\frac{2\ddot{a} h_1}{a}
-\frac{4\ddot{b} h_1}{b} -\frac{2 \dot{b}^2 h_1}{b^2} +\frac{h_1}{b^2} \left\lbrace  l(l+1)
\right\rbrace   \right] = -p_0 h_1 e \sin \theta (\partial_{\theta}Y), \label{11h} \\
\frac{e}{2} \left( \dot{h}_0+ \frac{\dot{a}h_0}{a} -\frac{h_1'}{a^2} \right) \left( \cos\theta (\partial_\theta Y)-\frac{2\sin \theta(\partial_\theta Y)}{\theta} +\sin\theta (\partial_\theta \partial_\theta Y) \right) =0. \label{11i}\hspace{2.5cm}
\end{eqnarray}

After simplifying and using the background field equations \eqref{2}, the above equations lead to the following relations:
\begin{eqnarray}
\Delta \cdot \rho_0 =0, \label{12a}  \hspace{3.5cm} \\
\Psi \cdot p_0 =0, \label{12b} \hspace{3.5cm}  \\
w (\rho_0 +p_0) =0, \label{12c} \hspace{3.2cm}  \\
v (\rho_0 +p_0) =0, \label{12d} \hspace{3.2cm}  \\
\frac{h_0''}{a^2} -\frac{\dot{h}_1'}{a^2}  +\frac{2 \dot{b}h_1'}{a^2 b} -\frac{h_0}{b^2} \left\lbrace  l(l+1) \right\rbrace  = 2U(\rho_0 +p_0), \label{12e} \hspace{0.5cm}  \\
\ddot{h}_1 -\dot{h}_0' -\frac{\dot{a}\dot{h}_1}{a}  +\frac{\dot{a}h_0'}{a} -\frac{2 \dot{b} h_0'}{b} -\frac{2\ddot{a} h_1}{a} +\frac{h_1}{b^2} \left\lbrace  l(l+1) \right\rbrace  = 0,  \label{12f}  \\
\dot{h}_0+ \frac{\dot{a}h_0}{a} -\frac{h_1'}{a^2} =0.  \hspace{3cm}   \label{12g}
\end{eqnarray}


From equations \eqref{12a} and \eqref{12b}, it can be concluded that $ \Delta =\Psi =0 $ even in the presence of matter when $ p_0 \neq 0 $, $ \rho_0 \neq 0 $. Also, $ w =v =0 $ if $ p_0 \neq - \rho_0 $, as deduced from equations \eqref{12c} and \eqref{12d}. Thus axial waves do not perturb the energy-density or pressure of the fluid. The only perturbation occurs in its azimuthal velocity $ U $ as evident from Eqn.\eqref{12e}.  Since the background solutions determine the matter content to be a stiff fluid \eqref{3e}, i.e. $ \rho_0 = p_0 \neq 0 $, their perturbations $ \Delta $ and $ \Pi $ will be equal, the r.h.s. of equations \eqref{11a} and \eqref{11b}, and hence equations \eqref{12a} and \eqref{12b} will be identical. We have assumed that $ \cot \theta =1/\theta $ in deriving the last terms of the equations \eqref{11g} and \eqref{11h}. This condition reduces to $ \tan \theta \simeq \theta $, which holds for small values of $ \theta $.

\subsection{Solutions to Perturbation equations in presence of matter: }

To solve the perturbation equation \eqref{12e}, we eliminate $ h_1(t,r) $ from it using Eqn.\eqref{12g}, and obtain
\begin{equation}\label{13a}
\frac{h_0''}{a^2} -\ddot{h}_0 -\dfrac{3\dot{a}}{a} \dot{h}_0 +\dfrac{2\dot{b}}{b} \dot{h}_0 -\dfrac{\dot{a}^2}{a^2}h_0 -\dfrac{\ddot{a}}{a}h_0 +\dfrac{2\dot{a}\dot{b}}{ab} h_0
-\frac{h_0}{b^2} \left\lbrace  l(l+1) \right\rbrace  = 2U(\rho_0 +p_0).
\end{equation}
Using Eqn.\eqref{4f} and equating $ p_0 $ to $ \rho_0 $, the above equation reads

\begin{equation}\label{13b}
\frac{h_0''}{b^4} -\ddot{h}_0 -\dfrac{4\dot{b}}{b} \dot{h}_0 - \dfrac{2\dot{b}^2}{b^2}h_0 - \dfrac{2\ddot{b}}{b}h_0
-\frac{h_0}{b^2} \left\lbrace  l(l+1) \right\rbrace  = 4U \rho_0.
\end{equation}
Now we write $ h_0(t,r) $ in terms of a new quantity $ \mathcal{Q}(t,r) $ as
\begin{equation}\label{13c}
h_0(t,r)= r^{\alpha} (b(t))^{\beta} \mathcal{Q}(t,r).
\end{equation}
Here $ \alpha $ and $ \beta $ can assume integral or fractional values. Substituting for $ b(t) $ and $ \rho_0(t) $ from Eqns.\eqref{4c} and \eqref{4f}, Eqn.\eqref{13b} in terms of $ \mathcal{Q}(t,r) $ becomes
\begin{equation}\label{13d}
\begin{split}
- \kappa ^\beta t^{\frac{\beta}{4}} r^\alpha \ddot{\mathcal{Q}} + \kappa ^{\beta -4} t^{\frac{\beta}{4} -1} r^\alpha \mathcal{Q}''  + \left[ \left( -\frac{1}{2}\beta -1 \right) \kappa ^\beta t^{(\frac{\beta}{4}-1)} r^\alpha \right]  \dot{\mathcal{Q}}
+ 2\alpha \kappa ^{\beta -4} t^{\frac{\beta}{4} -1} r^{\alpha -1}  \mathcal{Q}'  \\
+ \left[ \alpha^2 \kappa ^{\beta -4}  t^{\frac{\beta}{4} -1} r^{\alpha -2}  - \alpha \kappa ^{\beta -4} t^{\frac{\beta}{4} -1} r^{\alpha -2}
+  \left( - \frac{1}{16} \beta^2 +\frac{1}{4} \right) \kappa ^\beta t^{(\frac{\beta}{4}-2)} r^\alpha
-  l(l+1)  \kappa ^{\beta-2} t^{(\frac{\beta}{4}-\frac{1}{2})} r^\alpha \right] \mathcal{Q} = \dfrac{5}{4t^2} U.
\end{split}
\end{equation}
This is an inhomogeneous wave equation in $ \mathcal{Q}(t,r) $ with a source term $ U(t,r) $. This equation containing a single unknown variable $ \mathcal{Q}(t,r) $ may be termed as the \textbf{effective Regge-Wheeler equation} in Bianchi I background. It contains additional terms in first-order derivatives of $ \mathcal{Q}(t,r) $ when compared to the RW equation (Eqn.(87) in \cite{REZ}) for Schwarzschild black hole perturbations.
We have to solve this equation so as to determine the perturbation terms $ h_0 $ and $ h_1 $. This is the general perturbation equation containing the constants $ \kappa $ (from Eqn.\eqref{4c}), $ l $ (the angular momentum of spherical harmonics), $ \alpha $ and $ \beta $ (from Eqn.\eqref{13c}). In order to make the calculations tractable we need to insert particular values of the constants, and also assume a functional form of $ U(t,r) $. We choose to work with $ l=2 $ to get wave-like solutions \cite{SH1,SH2}. Putting $ \kappa=1 $, $ l=2 $, and $ \alpha = \beta =0 $, Eqn.\eqref{13d} reduces to the form
\begin{equation}\label{13e}
- \ddot{\mathcal{Q}} + \dfrac{1}{t} \mathcal{Q}'' - \dfrac{1}{t} \dot{\mathcal{Q}} + \dfrac{1}{4t^2} \mathcal{Q} -\dfrac{6}{t^{1/2}} \mathcal{Q} = \dfrac{5}{4t^2}U.
\end{equation}



\subsubsection*{Finding an expression for $ U $}

To solve \eqref{13e}, 
we require an explicit functional form of $ U $ in terms of $ t $ and $ r $. We can determine it from the normalisation condition \eqref{N} satisfied by the perturbed four-velocity. For the axially perturbed background \eqref{7b}, we find that the non-zero four-velocity components are:
\begin{eqnarray}
u^0 = g^{00}u_0 +g^{03}u_3 
= 1 +\dfrac{U}{h_0},   \hspace{0.6cm} 
u^3 = g^{33}u_3 + g^{30}u_0 = -\dfrac{eU \sin \theta (\partial_{\theta}Y)}{b^2 \theta ^2} +\dfrac{1}{e h_0 \sin \theta (\partial_{\theta}Y)} .
\end{eqnarray}
Therefore the normalisation condition becomes
\begin{equation}\label{14c}
\text{}  \hspace{0.3cm} u_{\mu}u^{\mu} = u_0u^0 + u_3u^3 
=1 +\frac{2U}{h_0} -\left( \dfrac{eU \sin \theta (\partial_{\theta}Y)}{b \theta}\right) ^2.  
\end{equation}
Comparing this equation \eqref{14c} with the general normalisation condition \eqref{N}, we arrive at the following two conditions.
\begin{equation}\label{14d}
\textrm{(I)} \quad 1 + \frac{2U}{h_0} \simeq 1, \qquad \text{and} \qquad
\textrm{(II)} \quad - \left( \dfrac{eU \sin \theta (\partial_{\theta}Y)}{b \theta}\right) ^2  \sim \mathcal{O}(e^2).
\end{equation}
Firstly, it can be concluded from the condition (I) that $ \dfrac{2U}{h_0} \ll 1 $, i.e. $ 2U \ll h_0 $. So the azimuthal four-velocity component arising due to axial waves will be much smaller in magnitude than the perturbation itself. Secondly, the term (containing $ e^2 $ and other higher powers of $ e $) on the right-hand side of the relation (II) has to be chosen in such a way that the functions of $ \theta $ cancel out from both sides, because $ U $ is a function of $ t $ and $ r $ only.
Considering a general expression of the form $  \mathcal{O}(e^2)= - \left( \dfrac{e \sin \theta (\partial_{\theta}Y)} {\theta} \right) ^2 (C T(t) R(r))^2  $, we obtain
\begin{equation}\label{14e}
 U(t,r) = \pm \hspace{0.1cm} Cb(t)T(t)R(r)  = \pm \hspace{0.1cm} \tilde{C} t^{1/4} TR.
\end{equation}
Here $ C $ is an arbitrary constant, $ T(t) $ and $ R(r) $ are some functions of $ t $ and $ r $, respectively. The functional form of $ b(t) $ is inserted from Eqn.\eqref{4f} and the product of the constants is represented by: $ \tilde{C} = C \kappa. $


\subsubsection*{Solutions in particular cases}

\begin{figure}[ht]
\centering
\begin{minipage}[b]{0.45\linewidth}
\includegraphics[height=4cm,width=8cm]{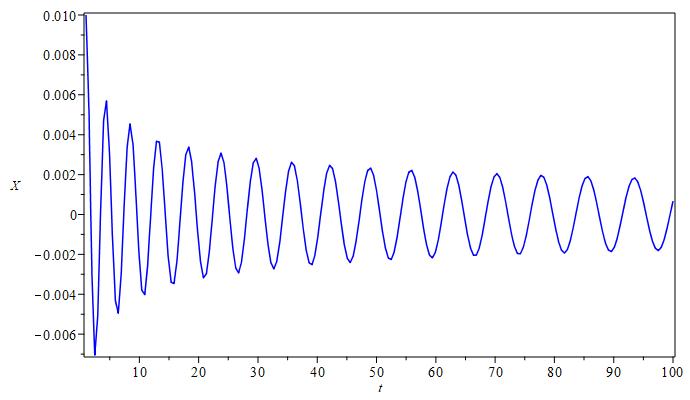}
\caption{$ X(t) $ is plotted against $ t $ .}  \label{fig:Plot1}  
\end{minipage}
\end{figure}

Finally, we are in the last step of finding the analytical solutions for the axial modes.
We have to plug in some particular value of $ \tilde{C} $, and some particular functions $ T(t) $ and $ R(r) $. Let us consider a particular case where $ C= \kappa=1 $, $ T(t)= t^{-\frac{1}{2}} $ and $ R(r)=r $ in Eqn.\eqref{14e}. Further, we take the positive value of $ U(t,r) $, and substitute it to solve Eqn.\eqref{13e}. The solution is obtained as:
%
%
%
%
%
\begin{equation}\label{15a}
\mathcal{Q}(t,r) = N/D,
\end{equation}
\begin{equation}\label{15b}
\begin{split}
\text{where} \qquad
N = 26208 \left[ \left( t^2 - \frac{\sqrt{t}}{32} \right) \text{HG} \left( [1], \left[\frac{1}{2}, \frac{7}{6} \right], -\frac{8}{3} t^{3/2} \right)
+ \left(\frac{256}{91} t^{7/2} \right) \text{HG} \left( [3], \left[\frac{5}{2}, \frac{19}{6} \right], -\frac{8}{3} t^{3/2} \right)
\right. \\ \left.
- \left( \frac{8}{7} t^2 \right) \text{HG} \left( [2], \left[\frac{3}{2}, \frac{13}{6} \right], -\frac{8}{3} t^{3/2} \right) \right]
 \sqrt{-t^{3/2}} \left( c_1 \exp(\sqrt{c_4} r /2) + c_2 \exp(- \sqrt{c_4} r /2) \right) c                                                                                                                                                                                                                                  _1 S(t)  \\
- 5460 \left[ \left( t^{1/4} \right) \text{HG} \left( [1], \left[\frac{1}{2}, \frac{7}{6} \right], -\frac{8}{3} t^{3/2} \right) c_1 \sqrt{-t^{3/2}}
+ \text{BesselI} \left(\frac{2}{3}, \frac{4}{3}\sqrt{6} \sqrt{-t^{3/2}} \right) c_2 t (-t^{3/2})^{1/6}
\right. \\ \left.
+ \text{BesselI} \left(\frac{4}{3}, \frac{4}{3}\sqrt{6} \sqrt{-t^{3/2}} \right) c_3 (-t^{3/2})^{5/6}
+ \left( \frac{\sqrt{6}}{12} \right) \text{BesselI} \left(\frac{1}{3}, \frac{4}{3}\sqrt{6} \sqrt{-t^{3/2}} \right) c_3  (-t^{3/2})^{1/3}  \right],
\\
\text{and} \qquad D = 26208 c_1 \sqrt{-t^{3/2}} \left[
\left( t^2 - \frac{\sqrt{t}}{32} \right) \text{HG} \left( [1], \left[\frac{1}{2}, \frac{7}{6} \right], -\frac{8}{3} t^{3/2} \right) + \left(\frac{256}{91} t^{7/2} \right) \text{HG} \left( [3], \left[ \frac{5}{2}, \frac{19}{6} \right], -\frac{8}{3} t^{3/2} \right)
\right. \\ \left.
- \left( \frac{8}{7} t^2 \right) \text{HG} \left( [2], \left[\frac{3}{2}, \frac{13}{6} \right], -\frac{8}{3} t^{3/2} \right)
\right]  .
\end{split}
\end{equation}
We used the Maple software to arrive at this analytical solution.
Here, $ c_1 $, $ c_2 $, $ c_3 $, and $ c_4 $ are integration constants, and HG denotes hypergeometric series. The BesselI functions are modified Bessel functions of the first kind.
However, in order to express \eqref{15b} in a compact form we have used an undetermined function $ S(t) $ to represent the solution of a second-order differential equation in an unknown function $ X(t) $ :
\begin{equation}\label{15i}
\ddot{X}(t) + \frac{1}{t} \dot{X}(t) - \frac{1}{4t}c_4 X(t) - \frac{1}{4t^2} X(t) + \frac{6}{\sqrt{t}} X(t) =0,
\end{equation}
that appeared in \eqref{15b} as obtained from the Maple software. Setting $ c_4 =1 $, the solution of Eqn.\eqref{15i} can be obtained by examining the nature of the plot in Fig.\eqref{fig:Plot1}. It is found to behave like a damped oscillator as time elapses. We may write an approximate trial solution: $ S (t) = (0.01) t^{1/3} (\sin t + \cos t) $ that will satisfy \eqref{15i}.

\subsubsection*{Discussions I}

\begin{figure}[h]
\centering
\includegraphics[height=8cm,width=14cm]{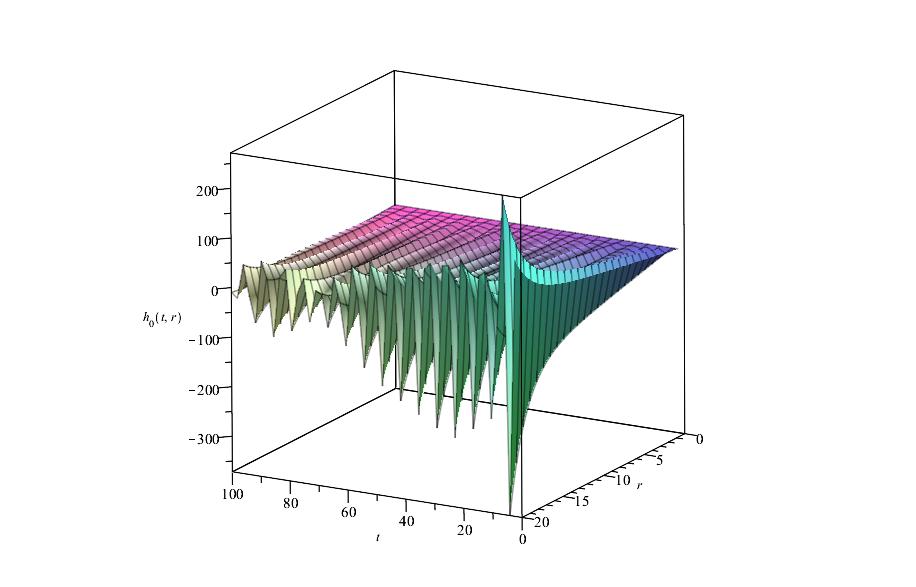}
\caption{The axial perturbation $ h_0 (t,r) $ is plotted against $ t $ and $ r $.}  
\label{fig:Plot3}
\end{figure}

The expression of $ \mathcal{Q} $ is found to be highly complicated. 
Equations \eqref{15a} and \eqref{15b} in a compact form read as:
\begin{equation} \label{16a}
\mathcal{Q}= [c_1 \exp(\sqrt{c_4} r /2) + c_2 \exp(- \sqrt{c_4} r /2)] S - \left( \frac{5 r}{24 c_1} \times \frac{\tilde{S_N}}{S_D} \right) ,
\end{equation}
%
\begin{eqnarray}
\begin{split}
\text{where} \qquad
\tilde{S}_N = \text{HG} \left( [1], \left[\frac{5}{6}, \frac{3}{2} \right], -\frac{8}{3} t^{\frac{3}{2}} \right) c_1 t^{\frac{1}{4}}
+ \text{BesselI} \left(\frac{2}{3}, \frac{4}{3}\sqrt{6} \sqrt{-t^{\frac{3}{2}}} \right) c_2 t (-t^{\frac{3}{2}})^{-\frac{1}{3}}   \hspace{2cm}  \label{16c} \\
+ \text{BesselI} \left(\frac{4}{3}, \frac{4}{3}\sqrt{6} \sqrt{-t^{\frac{3}{2}}} \right) c_3 (-t^{\frac{3}{2}})^{\frac{1}{3}}
+ \left( \frac{\sqrt{6}}{12} \right) \text{BesselI} \left(\frac{1}{3}, \frac{4}{3}\sqrt{6} \sqrt{-t^{\frac{3}{2}}} \right) c_3  (-t^{\frac{3}{2}})^{-\frac{1}{6}}  ,  \hspace{2cm} \\
S_D =
\left( t^2 - \frac{\sqrt{t}}{32} \right) \text{HG} \left( [1], \left[\frac{1}{2}, \frac{7}{6} \right], -\frac{8}{3} t^{\frac{3}{2}} \right) + \left(\frac{256}{91} t^{\frac{7}{2}} \right) \text{HG} \left( [3], \left[ \frac{5}{2}, \frac{19}{6} \right], -\frac{8}{3} t^{\frac{3}{2}} \right)
- \left( \frac{8}{7} t^2 \right) \text{HG} \left( [2], \left[\frac{3}{2}, \frac{13}{6} \right], -\frac{8}{3} t^{\frac{3}{2}} \right)
.  \label{16d}
\end{split}
\end{eqnarray}
Some of the terms occurring in the solutions of $ \mathcal{Q} $ contain imaginary parts. Only the real part of such terms have to be taken into account in order to get physically relevant solutions.
Given the expression for $ \mathcal{Q} (t,r) $, one can find $ h_0 (t,r) $ and subsequently $ h_1 (t,r) $.
From equations \eqref{13c} and \eqref{16a}, we get
\begin{equation}\label{17a}
h_0 (t,r) = r^0 (b(t))^0 \mathcal{Q} (t,r) =  \text{Re} \left[  \left( c_1 \exp \left( \frac{\sqrt{c_4} r}{2} \right)
+ c_2 \exp \left(- \frac{\sqrt{c_4} r}{2} \right) \right) S(t) - \left( \frac{5 r}{24 c_1} \times \frac{\tilde{S_N} (t)}{S_D (t)} \right)  \right].
\end{equation}
Setting all the constants to unity for the sake of simplicity, the 3-dimensional plot of $ h_0 (t,r) $ is obtained (using Maple software) as shown in Fig.\eqref{fig:Plot3}. We note that as mentioned in \cite{BR2}, for $ \mathcal{K} =0 $ the 2-dimensional hypersurfaces may be assumed to be open and infinite with the topology of $\mathcal{R}^2$. The coordinates $(r, \theta, \phi)$ are dimensionless, so that the scale factors have the dimension of length. The radial distance from the source is determined by the parameter $ra(t)$ and not by the parameter $r$ which have been plotted in this figure. Thus Fig.\eqref{fig:Plot3} clearly indicates that the perturbations $h_0 (t,r)$ decrease with time, although we have not been able to show the exact variation of the perturbations with increasing distance from the source.

Eqn.\eqref{12g} yields
\begin{equation}\label{17b}
\begin{split}
h_1 (t,r) = f_{sf} (t)+ \kappa^4 t \int_{r_0}^{r}\left( \dot{h}_0(t,r) + \frac{h_0(t,r)}{2t} \right) dr  \hspace{8cm} \\
=  \kappa^4 t \times  \text{Re}  \left[ \Lambda \left\lbrace \dot{S} (t) \left( c_1 e^{\Lambda (r- r_0)} - c_2 e^{- \Lambda (r-r_0)} \right) + \frac{{S} (t)}{2t} \left( c_1 e^{\Lambda (r-r_0)} + c_2 e^{- \Lambda (r-r_0)} \right) \right\rbrace \hspace{3cm}
\right. \\ \left.
- \frac{5}{24 c_1} \left\lbrace  \frac{d}{dt} \left( \frac{\tilde{S_N} (t)}{S_D (t)} \right) + \frac{1}{2t} \frac{\tilde{S_N} (t)}{S_D (t)} \right\rbrace
\right].   
\end{split}
\end{equation}
Here $ \Lambda = \sqrt{c_4} /2 $. In the particular case we have analysed here, all the constants in \eqref{17b} have been set to unity. The quantity $ f_{sf}(t) $ is the arbitrary integration constant that can be chosen to be zero as done in the last step. The parameter $ r_0 $ characterizes the initial hypersurface that generates the GWs. The suffix `$ sf $' is used to denote the case of stiff fluid.

Similar analysis can be done assuming other values of the constants $ \alpha $ and $ \beta $ (e.g. $ \alpha=0 $, $ \beta=1 $; $ \alpha=1 $, $ \beta=1 $; $ \alpha=1 $, $ \beta=0 $; $ \alpha=1 $, $ \beta=2 $), and for different powers of $ t $ in the function $ T(t) $ (e.g. 0, 1, 2, and -1/4) and those of $ r $ in $ R(r) $ (e.g. 0, and 1/2). We have found the solutions to be more or less similar to the one obtained here. The function $ T(t) $ has to be chosen keeping in mind that the condition (I) in \eqref{14d} remains valid.  


\subsection{In vacuum: Perturbation equations and Solutions}

As already mentioned earlier, this part has been studied in our previous work \cite{GD1}. The solution procedure is the same as above. Only the relation between the scale factors, $ a(t) $ and $ b(t) $ is defined by Eqn.\eqref{6} in the case of vacuum.

\section{Polar perturbations: Equations and Solutions}

We now discuss the polar perturbation equations and their analytical solutions for GWs propagating in Bianchi I universe. 

\subsection{In presence of matter: Perturbation equations}
We treat the cases $ \eta \neq 0 $ and $ \eta = 0 $ in separate paragraphs.

\subsection*{(i) Perturbation equations with $ \eta \neq 0 $}
In the presence of matter, the linearised field equations for the polar-perturbed metric \eqref{7f} are obtained as follows:

\begin{equation}\label{20a}
\begin{split}
(t-t) \hspace{0.1cm} \text{equation}:  \hspace{0.1cm}
\dfrac{4 \theta}{4[-b^2+ e\psi Y]^3 [-a^2+ eY \left\lbrace(1- a^2)(\chi+\psi) +2a^2 \eta \right\rbrace ]^2 \theta} \cdot \left[( a^4b^4\dot{b}^2
\right. \\ \left.
+ 2a^3\dot{a}b^5\dot{b}) +e \left\lbrace \frac{1}{2}a^2b^2(b^2\chi +b^2\psi +a^2\psi) \left( \frac{1}{\theta} (\partial_{\theta}Y) + (\partial_{\theta}\partial_{\theta}Y) \right) \right\rbrace
 \right. \\ \left.
+eY \left\lbrace ( 4a^3\dot{a}b^5\dot{b} - 2a\dot{a}b^5\dot{b} - 2a^2b^4\dot{b}^2 + 2a^4b^4\dot{b}^2)(\chi+\psi) - ( 4a^3\dot{a}b^3\dot{b} + a^4b^2\dot{b}^2)\psi
\right. \right. \\ \left. \left.
-2( 2a^4b^4\dot{b}^2 + 4a^3\dot{a}b^5\dot{b}) \eta - a^2b^5\dot{b}(\dot{\chi}+\dot{\psi}) - (a^4b^3\dot{b} + a^3\dot{a}b^4)\dot{\psi} +a^2b^4\psi'' +2a^2b^5\dot{b}\zeta'  \right\rbrace  \right]  \\
=  \rho_0 [1+ e(\Delta +\chi+\psi -2\eta) Y],
\end{split}
\end{equation}

\begin{equation}\label{20b}
\begin{split}
(r-r) \hspace{0.1cm}\text{equation}:  \hspace{0.1cm}
\dfrac{-4 \theta}{4[-b^2 + e\psi Y]^3 [-a^2+ eY \left\lbrace(1- a^2)(\chi+\psi) +2a^2 \eta \right\rbrace ]^2 \theta} \cdot \left[( - a^6b^4\dot{b}^2
\right. \\ \left.
- 2a^6b^5\ddot{b}) +e \left\lbrace \frac{1}{2}a^6b^2(b^2\chi +b^2\psi -\psi - 2b^2\eta) \left( \frac{1}{\theta} (\partial_{\theta}Y) + (\partial_{\theta}\partial_{\theta}Y) \right) \right\rbrace
 \right. \\ \left.
+eY \left\lbrace (- a^6b^4\dot{b}^2 + 3a^4b^4\dot{b}^2 + 6a^4b^5\ddot{b} - 2a^6b^5\ddot{b} )(\chi+\psi) + (4a^6b^3\ddot{b} + 3a^6b^2\dot{b}^2)\psi
\right. \right. \\ \left. \left.
+ 2 ( a^6b^4\dot{b}^2
+ 2a^6b^5\ddot{b}) \eta + a^6b^5\dot{b}(\dot{\chi}+\dot{\psi} -2\dot{\eta}) - a^6b^3\dot{b}\dot{\psi} + a^6b^4\ddot{\psi}  \right\rbrace  \right] \\
= p_0 [a^2 + e(a^2 \Pi -\chi-\psi)Y],
\end{split}
\end{equation}

\begin{equation}\label{20c}
\begin{split}
(\theta-\theta) \hspace{0.1cm}\text{equation}:  \hspace{0.1cm}
\dfrac{-4 \theta}{4[-b^2 +e\psi Y] [-a^2+ eY \left\lbrace(1- a^2)(\chi+\psi) +2a^2 \eta \right\rbrace ]^2 \theta} \cdot \left[( - a^4b^3\ddot{b} -a^3\ddot{a}b^4
\right. \\ \left.
- a^3\dot{a}b^3\dot{b}) +eY \left\lbrace \frac{1}{2Y} ((a^4b^2 -a^2b^2)(\chi+\psi) -2a^4b^2 \eta) (\frac{1}{\theta} (\partial_{\theta}Y))
+ (a\ddot{a}b^4 - a^4b^3\ddot{b} - a^3\ddot{a}b^4
\right. \right. \\ \left. \left.
- a^3\dot{a}b^3\dot{b} + a\dot{a}b^3\dot{b} + 2a^2b^3\ddot{b} + \dot{a}^2b^4)(\chi+\psi) + (a^4\dot{b}^2 + a^4b\ddot{b} + a^3\dot{a}b\dot{b} + 2a^3\ddot{a}b^2)\psi
\right. \right. \\ \left. \left.
+ 2(a^4b^3\ddot{b} + a^3\ddot{a}b^4 + a^3\dot{a}b^3\dot{b}) \eta
+ (-a\dot{a}b^4 + \frac{1}{2}a^3\dot{a}b^4 + \frac{1}{2}a^4b^3\dot{b} + \frac{1}{2}a^2b^3\dot{b})(\dot{\chi}+\dot{\psi})
\right. \right. \\ \left. \left.
+ ( - a^4b\dot{b} + \frac{1}{2}a^3\dot{a}b^2)\dot{\psi} - (a^4b^3 \dot{b} + a^3 \dot{a} b^4) \dot{\eta} - a^2b^3\dot{b}\zeta'
 + \frac{1}{2} a^2b^4(\ddot{\chi}+\ddot{\psi}+\chi''+\psi'')
\right. \right. \\ \left. \left.
 + \frac{1}{2}a^4b^2 \ddot{\psi} - \frac{1}{2}a^2b^2 \psi'' - a^2b^4 \eta'' - a^2b^4\dot{\zeta}' \right\rbrace \right]
=  p_0 [b^2 + e(b^2 \Pi -\psi)Y],
\end{split}
\end{equation}

\begin{equation}\label{20d}
\begin{split}
(\phi-\phi) \hspace{0.1cm}\text{equation}:  \hspace{0.1cm}
\dfrac{-4 \theta ^2}{4[-b^2 +e\psi Y] [-a^2+ eY \left\lbrace(1- a^2)(\chi+\psi)]^2 +2a^2 \eta \right\rbrace } \cdot \left[( - a^4b^3\ddot{b}-a^3\ddot{a}b^4
\right. \\ \left.
- a^3\dot{a}b^3\dot{b}) +eY \left\lbrace \frac{1}{2Y} ((a^4b^2 -a^2b^2)(\chi+\psi) -2a^4b^2 \eta) (\partial_{\theta}\partial_{\theta}Y)
+ (a\ddot{a}b^4 - a^4b^3\ddot{b} - a^3\ddot{a}b^4
\right. \right. \\ \left. \left.
- a^3\dot{a}b^3\dot{b} + a\dot{a}b^3\dot{b} + 2a^2b^3\ddot{b} + \dot{a}^2b^4)(\chi+\psi) + (a^4\dot{b}^2 + a^4b\ddot{b} + a^3\dot{a}b\dot{b} + 2a^3\ddot{a}b^2) \psi
\right. \right. \\ \left. \left.
+ 2(a^4b^3\ddot{b} + a^3\ddot{a}b^4 + a^3\dot{a}b^3\dot{b}) \eta
+ (-a\dot{a}b^4 + \frac{1}{2}a^3\dot{a}b^4 + \frac{1}{2}a^4b^3\dot{b} + \frac{1}{2}a^2b^3\dot{b}) (\dot{\chi}+\dot{\psi})
\right. \right. \\ \left. \left.
+ ( - a^4b\dot{b} + \frac{1}{2}a^3\dot{a}b^2)\dot{\psi} - (a^4b^3 \dot{b} + a^3 \dot{a} b^4) \dot{\eta} - a^2b^3\dot{b}\zeta'
+ \frac{1}{2} a^2b^4(\ddot{\chi}+\ddot{\psi}+\chi''+\psi'')
\right. \right. \\ \left. \left.
+ \frac{1}{2}a^4b^2 \ddot{\psi} - \frac{1}{2}a^2b^2 \psi'' - a^2b^4 \eta'' - a^2b^4\dot{\zeta}'  \right\rbrace \right]
=  p_0 [b^2 + e(b^2 \Pi -\psi)Y] \theta ^2,
\end{split}
\end{equation}

\begin{equation}\label{20e}
\begin{split}
(t-r) \hspace{0.1cm} \text{equation}:  \hspace{0.3cm}
\dfrac{ea^4b^4 \theta}{[-b^2+ e\psi Y]^3 [a^2- eY \left\lbrace (1- a^2)(\chi+\psi) +2a^2 \eta \right\rbrace ]^2 \theta} \cdot \left[ \left\lbrace  -\dot{\psi}' +\left(\frac{\dot{a}}{a} + \frac{\dot{b}}{b}\right) \psi'
\right. \right. \\ \left. \left.
- b\dot{b}(\chi'+\psi' -2\eta') -(2b\ddot{b} + \dot{b}^2) \zeta \right\rbrace Y
-\frac{\zeta}{2} \left\lbrace (\partial_{\theta} \partial_{\theta} Y) + \frac{1}{\theta} (\partial_{\theta} Y)\right\rbrace  \right]
=  e[(\rho_0 +p_0)aw -p_0\zeta]Y,
\end{split}
\end{equation}

\begin{equation}\label{20f}
\begin{split}
(t-\theta) \hspace{0.1cm} \text{equation}:  \hspace{0.2cm}
\dfrac{e ab(\partial_{\theta} Y)}{2[-b^2 + e\psi Y]^2 [a^2- eY \left\lbrace(1- a^2)(\chi+\psi) + 2a^2 \eta \right\rbrace ]^2} \cdot \left[(-\dot{a}b^3 - ab^2\dot{b} + a^2\dot{a}b^3
 \right. \\ \left.
+ a^3b^2 \dot{b}) (\chi+\psi) - 2a^3 \dot{b} \psi - 2(a^2\dot{a}b^3 + a^3b^2\dot{b}) \eta + ab^3 (\dot{\chi} +\dot{\psi} -\zeta') + a^3b \dot{\psi} \right]
= e(\rho_0 +p_0) v(\partial_{\theta}Y),
\end{split}
\end{equation}

\begin{equation}\label{20g}
\begin{split}
(r-\theta) \hspace{0.1cm} \text{equation}:  \hspace{0.2cm}
\dfrac{e (\partial_{\theta} Y) a^4b^4}{2[-b^2+ e\psi Y]^2 [a^2- eY \left\lbrace(1- a^2)(\chi+\psi) +2a^2 \eta \right\rbrace ]^2} \cdot \left[-(\chi' +\psi' -2\eta' -\dot{\zeta})
+ \frac{1}{b^2}\psi' +\frac{\dot{a}}{a}\zeta  \right] =0,
\end{split}
\end{equation}

\begin{equation}\label{20h}
(t-\phi) \hspace{0.3cm} \text{equation}:  \hspace{0.3cm}
e(\rho_0 +p_0) U \sin \theta (\partial_{\theta}Y) =0.
\end{equation}

The assumption: $ \cot \theta =1/\theta $ is applied to Eqns.\eqref{20a}-\eqref{20e} also. As in the axial perturbation equations, this condition implies that $ \tan \theta \simeq \theta $, hence $ \theta $ must be small. Neglecting all terms containing second or higher orders of $ e $ in the expansions appearing here, and using the background field equations \eqref{2}, and the relation \eqref{sph}, the above set of equations lead to the following:

\begin{equation}\label{21a}
w= \frac{1}{ab^2 (\rho_0+p_0)} \left[ - \dfrac{l(l+1)}{2} \zeta + b\dot{b} \chi' +\left(-\frac{\dot{a}}{a} - \frac{\dot{b}}{b} + b\dot{b}\right) \psi' -2b \dot{b} \eta' + \dot{\psi}' \right],
\end{equation}

\begin{equation}\label{21b}
\begin{split}
v= \frac{1}{2(\rho_0+p_0)} \left[ \left( -\frac{\dot{a}}{a^3} + \frac{\dot{a}}{a} - \frac{\dot{b}}{a^2b} + \frac{\dot{b}}{b} \right)\chi
+\left( -\frac{\dot{a}}{a^3} + \frac{\dot{a}}{a} - \frac{\dot{b}}{a^2b} + \frac{\dot{b}}{b} - \frac{2\dot{b}}{b^3}\right)\psi
\right. \\ \left.
- 2\left( \frac{\dot{a}}{a} +   \frac{\dot{b}}{b}\right) \eta + \frac{1}{a^2}\dot{\chi} + \left( \frac{1}{a^2}+\frac{1}{b^2}\right) \dot{\psi} - \frac{1}{a^2}\zeta' \right],
\end{split}
\end{equation}
\begin{equation}\label{21c}
U =0,
\end{equation}

\begin{equation}\label{21d}
\frac{\dot{a}}{a}\zeta +\dot{\zeta} -\chi' +\left(\frac{1}{b^2}-1\right) \psi' +2\eta' =0,
\end{equation}

\begin{equation}\label{21e}
\begin{split}
\Delta = \frac{1}{a^4b^6\rho_0} \left[ \left\lbrace 2a\dot{a}b^5\dot{b} - 2a^3\dot{a}b^5\dot{b}  -a^4b^4\dot{b}^2
-\frac{l(l+1)}{2} \cdot a^2b^4 \right\rbrace \chi
\right. \\ \left.
 + \left\lbrace 2a\dot{a}b^5\dot{b} -2a^3\dot{a}b^5\dot{b} + 2a^3\dot{a}b^3\dot{b} + 2a^4b^2\dot{b}^2 - a^4b^4\dot{b}^2 -\frac{l(l+1)}{2} \cdot a^2b^2(a^2+b^2) \right\rbrace \psi
\right. \\ \left.
+ 2(2a^3\dot{a} b^5 \dot{b} + a^4b^4 \dot{b}^2) \eta - a^2b^5\dot{b}\dot{\chi} - \left\lbrace a^2b^5\dot{b} + a^4b^3\dot{b} + a^3\dot{a}b^4 \right\rbrace \dot{\psi}
+ a^2b^4\psi'' +2a^2b^5\dot{b}\zeta'  \right],
\end{split}
\end{equation}

\begin{equation}\label{21f}
\begin{split}
\Pi = \frac{1}{a^6b^6p_0} \left[ \left\lbrace a^6b^4\dot{b}^2 + 2a^6b^5\ddot{b} - \frac{l(l+1)}{2} \cdot a^6b^4 \right\rbrace \chi + \left\lbrace a^6b^4\dot{b}^2 - 2a^6b^3\ddot{b} + 2a^6b^5\ddot{b}
\right. \right. \\ \left. \left.
 - \frac{l(l+1)}{2} \cdot a^6b^2(b^2 -1)\right\rbrace \psi
+ 2  \left\lbrace - a^6b^4 \dot{b}^2 - 2 a^6b^5 \ddot{b} + \frac{l(l+1)}{2} \cdot a^6b^4 \right\rbrace \eta
\right. \\ \left.
 + a^6b^5\dot{b}\dot{\chi} + (a^6b^5\dot{b} - a^6b^3\dot{b})\dot{\psi} - 2a^6b^5\dot{b}\dot{\eta} + a^6b^4\ddot{\psi}
 \right],
\end{split}
\end{equation}

\begin{equation}\label{21g}
\begin{split}
p_0 b^2\Pi  = \frac{1}{a^4b^2} \left[ ( \dot{a}^2b^4 - a\ddot{a}b^4 + a^4b^3\ddot{b} + a^3\ddot{a}b^4 + a^3\dot{a}b^3\dot{b}
- a\dot{a}b^3\dot{b} )(\chi+\psi) + (a^4\dot{b}^2 - a^4 b\ddot{b}
 \right. \\ \left.
-  a^3\dot{a}b\dot{b})\psi - 2 (a^4b^3\ddot{b} + a^3\ddot{a} b^4 + a^3\dot{a} b^3\dot{b}) \eta
+ (-a\dot{a}b^4 + \frac{1}{2} a^3\dot{a}b^4 + \frac{1}{2} a^4b^3\dot{b} + \frac{1}{2} a^2b^3\dot{b}) (\dot{\chi} +\dot{\psi})
\right. \\ \left.
+ (- a^4b\dot{b} + \frac{1}{2}a^3\dot{a}b^2) \dot{\psi}
- (a^4b^3 \dot{b} + a^3 \dot{a} b^4) \dot{\eta}  - a^2b^3\dot{b}\zeta'
+ \frac{1}{2} a^2b^4 (\ddot{\chi} +\ddot{\psi} +\chi'' +\psi'')
\right. \\ \left.
+ \frac{1}{2}a^4b^2 \ddot{\psi} - \frac{1}{2}a^2b^2 \psi'' - a^2b^4 \eta'' - a^2b^4\dot{\zeta}'
+ \frac{1}{2Y} \left\lbrace a^2b^2(a^2 -1)(\chi+\psi) -2a^4b^2 \eta \right\rbrace \frac{1}{\theta} (\partial_{\theta}Y)  \right],
\end{split}
\end{equation}

\begin{equation}\label{21h}
\begin{split}
p_0 b^2\Pi  = \frac{1}{a^4b^2} \left[ (\dot{a}^2b^4 - a\ddot{a}b^4 + a^4b^3\ddot{b} + a^3\ddot{a}b^4 + a^3\dot{a}b^3\dot{b}
- a\dot{a}b^3\dot{b})(\chi+\psi) + (a^4\dot{b}^2 - a^4 b\ddot{b}
\right. \\ \left.
- a^3\dot{a}b\dot{b})\psi - 2(a^4b^3\ddot{b} + a^3\ddot{a}b^4 + a^3\dot{a}b^3\dot{b}) \eta
+ (-a\dot{a}b^4 + \frac{1}{2} a^3\dot{a}b^4 + \frac{1}{2} a^4b^3\dot{b} + \frac{1}{2} a^2b^3\dot{b}) (\dot{\chi} +\dot{\psi})
\right. \\ \left.
+ (- a^4b\dot{b} + \frac{1}{2}a^3\dot{a}b^2) \dot{\psi}
- (a^4b^3 \dot{b} + a^3 \dot{a} b^4) \dot{\eta}  - a^2b^3\dot{b}\zeta'
+ \frac{1}{2} a^2b^4 (\ddot{\chi} +\ddot{\psi} +\chi'' +\psi'')
\right. \\ \left.
+ \frac{1}{2}a^4b^2 \ddot{\psi} - \frac{1}{2}a^2b^2 \psi'' - a^2b^4 \eta'' - a^2b^4\dot{\zeta}'
+ \frac{1}{2Y} \left\lbrace a^2b^2(a^2 -1)(\chi+\psi) -2a^4b^2 \eta \right\rbrace (\partial_{\theta} \partial_{\theta}Y) \right].
\end{split}
\end{equation}

Adding Eqns.\eqref{21g} and \eqref{21h} and substituting Eqn.\eqref{21f}, we arrive at
\begin{equation}\label{21i}
\begin{split}
\left[ \dot{a}^2b^4 - a\ddot{a}b^4 - a^4b^3\ddot{b} + a^3\ddot{a}b^4 + a^3\dot{a}b^3\dot{b} - a\dot{a}b^3\dot{b} -a^4b^2\dot{b}^2
+ \frac{l(l+1)}{4} (a^4b^2 +a^2b^2) \right] \chi \\
+ \left[ -a\ddot{a}b^4 - a^4b^3\ddot{b} + a^3\ddot{a}b^4 + a^3\dot{a}b^3\dot{b} - a\dot{a}b^3\dot{b} - a^3\dot{a}b\dot{b} + \dot{a}^2b^4 + a^4\dot{b}^2 + a^4 b\ddot{b} - a^4b^2\dot{b}^2
\right. \\ \left.
+ \frac{l(l+1)}{4} (a^4b^2 +a^2b^2 -2a^4) \right]  \psi
+ \left[ 2a^4b^3\ddot{b} - 2a^3\ddot{a}b^4 - 2a^3\dot{a}b^3\dot{b} + 2a^4b^2 \dot{b}^2 - \frac{l(l+1)}{2} \cdot a^4b^2 \right] \eta \\
+ \left[ \frac{1}{2}a^3\dot{a}b^4 - \frac{1}{2}a^4b^3\dot{b} + \frac{1}{2}a^2b^3\dot{b} - a\dot{a}b^4 \right] \dot{\chi}
+ \left[ \frac{1}{2}a^3\dot{a}b^4 - \frac{1}{2}a^4b^3\dot{b} + \frac{1}{2}a^2b^3\dot{b} + \frac{1}{2}a^3\dot{a}b^2 -a\dot{a}b^4 \right] \dot{\psi}\\ + (a^4b^3 \dot{b} - a^3 \dot{a} b^4) \dot{\eta}  - a^2b^3\dot{b}\zeta'
 + \frac{1}{2} a^2b^4 \left(\ddot{\chi} +\chi''\right)  \\
+ \left(\frac{1}{2}a^2b^4 - \frac{1}{2}a^4b^2\right) \ddot{\psi} + \left(\frac{1}{2}a^2b^4 - \frac{1}{2}a^2b^2\right) \psi'' - a^2b^4 \eta'' - a^2b^4\dot{\zeta}' =0 .
\end{split}
\end{equation}

Equations \eqref{21a} and \eqref{21b} determine the perturbation of the two components of the fluid four-velocity. From Eqn.\eqref{21c}, one finds that the third component, i.e. the azimuthal velocity $ U $ is zero in case of polar waves. These properties distinguish polar GWs from axial GWs. Moreover, the polar perturbations are marked in the energy density and pressure as evident from Eqns.\eqref{21e} and \eqref{21f} respectively. Eqns.\eqref{21d} and \eqref{21i} do not explicitly depend on matter. Eqns.\eqref{21a}, \eqref{21b} and \eqref{21e} are the constraint equations. Following Clarkson \textit{et al.} \cite{CL1}, we can say that Eqns.\eqref{21d}, \eqref{21f} and \eqref{21i}, together with the constraint equation $ \eta =0 $ are the three master equations governing the evolution of the perturbing variables. Also, if we suppose that all the matter perturbations become zero, then inserting $ \Pi = \Delta =w =v =0 $ in Eqns.\eqref{21a}, \eqref{21b}, \eqref{21e} and \eqref{21f} yields $ \chi = \psi = \zeta =0 $, i.e. the polar perturbations vanish. Hence it can be concluded that polar GWs must bring about matter inhomogeneities and anisotropies in the background they travel through. As one can see from the set of equations \eqref{21a}-\eqref{21i}, the polar perturbation variables are heavily coupled to one another. Therefore, the Zerilli equation in a single variable cannot be extracted. However, we may identify the Zerilli equation from \eqref{21i} under suitable assumptions. We elaborate this point in \textit{Discussions II} while analysing Case 3 of the solutions for vacuum perturbations.

\subsection*{(ii) Perturbation equations with $ \eta = 0 $}
When $ \eta =0 $, the perturbation equations \eqref{21a}-\eqref{21f} and \eqref{21i} become  a little simpler. We will use this constraint to solve the equations analytically.

\begin{center}
\textbf{ \underline{The particular case of stiff fluid} }
\end{center}
We now concentrate on the stiff fluid as the matter content of the Bianchi I universe. The fluid pressure and energy density being equal, the polar perturbation equations with both zero or non-zero value of $ \eta $ are slightly more simplified.
\subsection*{(a) Perturbation equations with $ \eta \neq 0 $}
The polar-perturbed equations \eqref{21a}-\eqref{21e} in the presence of a stiff fluid get reduced to:

\begin{equation}\label{23a}
w= \frac{2}{5 \kappa^4} \left[ - 2t l(l+1) \zeta + \kappa^2 t^{1/2} \chi' + \left( \kappa^2 t^{1/2} - 3 \right) \psi' - 2\kappa^2 t^{1/2} \eta'
+ 4t \dot{\psi}' \right]
\end{equation}

\begin{equation}\label{23b}
v= \frac{1}{5 \kappa^4} \left[ ( 3\kappa^4 t - 3 )\chi  + ( 3\kappa^4 t - 3 - 2\kappa^2 t^{1/2} ) \psi
 - 6\kappa^4 t \eta + 4t \dot{\chi} + ( 4t + 4\kappa^2 t^{3/2} ) \dot{\psi} - 4t \zeta' \right],
\end{equation}

\begin{equation}\label{23c}
U =0,
\end{equation}

\begin{equation}\label{23d}
\frac{1}{2t}\zeta + \dot{\zeta} - \chi' + \left( \frac{1}{\kappa^2 t^{1/2}} -1 \right) \psi' + 2\eta' =0,
\end{equation}

\begin{equation}\label{23e}
\begin{split}
\Delta = \frac{2}{5 \kappa^4} \left[ \left\lbrace \frac{2}{t} - \frac{5 \kappa^4}{2} - \frac{4 t^{1/2}}{\kappa^2} l(l+1) \right\rbrace \chi
+ \left\lbrace \frac{2}{t} + \frac{3\kappa^2}{t^{1/2}} - \frac{5 \kappa^4}{2} - 4 \left( \frac{t^{1/2}}{\kappa^2} + t \right) l(l+1) \right\rbrace \psi - 5 \kappa^4 \eta
\right. \\ \left.
- 2 \dot{\chi} - \left\lbrace 2 + 6 \kappa^2 t^{1/2} \right\rbrace \dot{\psi} + \frac{8t^{1/2}}{\kappa^2} \psi'' + 4 \zeta'  \right]
\end{split}
\end{equation}

Since $ p_0 = \rho_0 $, their respective perturbations $ \Delta $ and $ \Pi $ must be equated. Hence, inserting $ a=b^2 $ in Eqns.\eqref{21f} and \eqref{21i}, replacing $ \Pi $ by $ \Delta $ and plugging in the expression for $ \Delta $ from Eqn.\eqref{23e}, we get respectively:
\begin{equation}\label{23f}
\begin{split}
 \frac{4}{5 \kappa^4} \left[ \left\lbrace  \frac{1}{t} + (2 \kappa^2 t^{3/2} - \frac{2 t^{1/2}}{\kappa^2} ) l(l+1)  \right\rbrace \chi
+ \left\lbrace  \frac{1}{t}
+ ( 2 \kappa^2 t^{3/2} - 4t - \frac{2 t^{1/2}}{\kappa^2} ) l(l+1) \right\rbrace \psi
- 4 \kappa^2 t^{3/2} l(l+1) \eta
\right. \\ \left.
 - (1 + \kappa^4 t)  \dot{\chi} - \left( 1 + \kappa^4 t + 2 \kappa^2 t^{1/2} \right) \dot{\psi}
+ 2\kappa^4 t \dot{\eta} +  2 \zeta' - 4 \kappa^2 t^{3/2} \ddot{\psi} + \frac{4 t^{1/2}}{\kappa^2} \psi''
\right]  =0,
\end{split}
\end{equation}

\begin{equation}\label{23g}
\begin{split}
\frac{2}{5 \kappa^4} \left[ \left\lbrace - \frac{1}{t} + \left( 2 \kappa^2 t^{3/2} - \frac{6 t^{1/2}}{\kappa^2} \right) l(l+1) \right\rbrace \chi
+  \left\lbrace \frac{2 \kappa^2}{t^{1/2}} - \frac{1}{t} + ( 2 \kappa^2 t^{3/2} - \frac{6 t^{1/2}}{\kappa^2} - 4t ) l(l+1) \right\rbrace \psi
\right. \\ \left.
-  4 \kappa^2 t^{3/2} l(l+1) \eta
+ \left\lbrace 1 - 3 \kappa^4 t \right\rbrace \dot{\chi} + \left\lbrace 1 - 3 \kappa^4 t - 6 \kappa^2 t^{1/2} \right\rbrace \dot{\psi}
+ 6 \kappa^4 t \dot{\eta} + 6 \zeta'
\right. \\ \left.
 - 4t \ddot{\chi} + \left\lbrace - 4 \kappa^2 t^{3/2} - 4t \right\rbrace \ddot{\psi} - 4t \chi'' + \left( \frac{12 t^{1/2}}{\kappa^2} - 4t \right) \psi''
+ 8t \eta'' + 8t \dot{\zeta}'
\right] =0.
\end{split}
\end{equation}

Taking the time-derivatives of equations \eqref{23a}-\eqref{23c}, the evolution of $ w $, $ v $ and $ \Delta $ are found to be governed by the following equations :
\begin{equation}\label{23h}
\begin{split}
\dot{w} = \frac{2}{5 \kappa^4} \left[ -2 l(l+1) \zeta - 2t l(l+1) \dot{\zeta} + \frac{\kappa^2}{2 t^{1/2}} \chi'
 + \frac{\kappa^2}{2 t^{1/2}} \psi' - \frac{\kappa^2}{ t^{1/2}} \eta' +  \kappa^2 t^{1/2} \dot{\chi}'
+ ( \kappa^2 t^{1/2} + 1 ) \dot{\psi}'
 - 2 \kappa^2 t^{1/2} \dot{\eta}' + 4t \ddot{\psi}' \right],
\end{split}
\end{equation}
\begin{equation}\label{23i}
\begin{split}
\dot{v} = \frac{1}{5 \kappa^4} \left[ 3 \kappa^4 \chi + \left(3 \kappa^4 - \frac{\kappa^2}{t^{1/2}} \right) \psi - 6 \kappa^4 \eta  - 4\zeta' + (3 \kappa^4 t + 1) \dot{\chi}
 + \left( 3 \kappa^4 t + 1 + 4 \kappa^2 t^{1/2} \right) \dot{\psi} - 6 \kappa^4 t \dot{\eta} + 4t \ddot{\chi}
 \right. \\ \left.
 + (4t +4 \kappa^2 t^{3/2}) \ddot{\psi} - 4t \dot{\zeta}'
\right],
\end{split}
\end{equation}
\begin{equation}\label{23j}
\begin{split}
\dot{\Delta} = \frac{4}{5 \kappa^4} \left[ \left\lbrace - \frac{1}{t^2} - \frac{1}{\kappa^2 t^{1/2}} l(l+1)  \right\rbrace  \chi
+ \left\lbrace - \frac{1}{t^2} - \frac{3 \kappa^2}{4 t^{3/2}} - \left(2 + \frac{1}{\kappa^2 t^{1/2}} \right) l(l+1) \right\rbrace  \psi
\right. \\ \left.
+ \left\lbrace \frac{1}{t} - \frac{5 \kappa^4}{4}
- \frac{2 t^{1/2}}{\kappa^2} l(l+1) \right\rbrace \dot{\chi}
+ \left\lbrace  \frac{1}{t} - \frac{5 \kappa^4}{4} - \left( 2t + \frac{2 t^{1/2}}{\kappa^2} \right) l(l+1) \right\rbrace  \dot{\psi}
\right. \\ \left.
 + \frac{5 \kappa^4}{2} \dot{\eta} - \ddot{\chi} - (1 + 3 \kappa^2 t^{1/2}) \ddot{\psi} + \frac{2}{\kappa^2 t^{1/2}} \psi'' + 2\dot{\zeta}'
+ \frac{4 t^{1/2}}{\kappa^2} \dot{\psi}''
\right].
\end{split}
\end{equation}

Following \cite{CL1}, it can be said that for polar perturbations to Bianchi I background, equations \eqref{23b}, \eqref{23d}, \eqref{23f}, \eqref{23g} and \eqref{23i} hold for $ l \geq 1 $, and the rest, \eqref{23a}, \eqref{23e}, \eqref{23h} and \eqref{23j}, for $ l \geq 0 $.

\subsection*{(b) Perturbation equations with $ \eta= 0 $}
For $ l \geqslant 2 $, the above equations will hold with the terms containing $ \eta $ set to zero.

\subsection{In presence of matter: Solutions to Perturbation equations}
As evident from the set of polar perturbation equations, in contrast to the axial perturbation equations, the polar solutions cannot be derived straight away. Eqn.\eqref{23d} containing the three perturbation elements does not help in simplifying the remaining equations and extracting a differential equation in a single variable. The above equations are highly complicated even when $ \eta = 0 $, and cannot be solved analytically without certain simplifying assumptions. We attempt to solve equation \eqref{23g}. Choosing $ \eta = 0 $, $ l=2 $, $ \kappa =1 $, and assuming $ \psi (t,r) = 0 $, and $ \zeta (t,r) =q_{sf} \chi (t,r) $, $ q_{sf} $ being an arbitrary constant, this equation reduces to
\begin{equation}\label{24a}
\left( 12t^{3/2} - 12t^{1/2} + \frac{1}{t} \right) \chi - (1+t) \dot{\chi} + 2q_{sf} \chi' =0.
\end{equation}
The corresponding solution is obtained as:
\begin{equation}\label{24b}
\chi (t,r) = \mathcal{F} (r+ 2q_{sf} \ln (1+t)) \left( \frac{t}{1+t} \right) \exp \left\lbrace 8t^{3/2} + 48 \arctan t^{1/2} - 48t^{1/2} \right\rbrace.
\end{equation}
Here, $ \mathcal{F} $ is an undetermined function of both $ t $ and $ r $. But this does not conform to the Regge-Wheeler scheme where the function of $ t $ and that of $ r $ must separate out as product in the solutions of the perturbation equations.

\subsection{In vacuum: Perturbation equations}
In the absence of matter, the set of polar perturbation equations \eqref{21a}-\eqref{21i} with $ \eta= 0 $ are reduced to the following:
\begin{equation}\label{25a}
- \dfrac{l(l+1)}{2} \zeta + b\dot{b} \chi' +\left(-\frac{\dot{a}}{a} - \frac{\dot{b}}{b} + b\dot{b}\right) \psi' + \dot{\psi}' =0,
\end{equation}

\begin{equation}\label{25b}
\begin{split}
\left( -\frac{\dot{a}}{a^3} + \frac{\dot{a}}{a} - \frac{\dot{b}}{a^2b} + \frac{\dot{b}}{b} \right)\chi +\frac{1}{a^2}\dot{\chi}
+\left( -\frac{\dot{a}}{a^3} + \frac{\dot{a}}{a} - \frac{\dot{b}}{a^2b} + \frac{\dot{b}}{b} - \frac{2\dot{b}}{b^3}\right)\psi
+ \left( \frac{1}{a^2}+\frac{1}{b^2}\right) \dot{\psi} - \frac{1}{a^2}\zeta'  =0,
\end{split}
\end{equation}

\begin{equation}\label{25c}
\frac{\dot{a}}{a}\zeta +\dot{\zeta}-\chi' +\left(\frac{1}{b^2}-1\right) \psi' =0,
\end{equation}

\begin{equation}\label{25d}
\begin{split}
\left\lbrace \frac{2\dot{a}b\dot{b}}{a} -\frac{l(l+1)}{2} \right\rbrace \chi + \left\lbrace \frac{2\dot{a}b\dot{b}}{a} - \frac{2a \dot{a} \dot{b}}{b} -\frac{l(l+1)}{2} \left( 1+\frac{a^2}{b^2} \right) \right\rbrace \psi - b\dot{b}\dot{\chi}
- \left\lbrace b\dot{b} + a\dot{a} + \frac{a^2\dot{b}}{b} \right\rbrace \dot{\psi}
 + \psi'' + 2b\dot{b}\zeta'  =0,
\end{split}
\end{equation}

\begin{equation}\label{25e}
\left\lbrace - \frac{l(l+1)}{2} \right\rbrace \chi  + \left\lbrace \frac{\dot{b}^2}{b^2} - \frac{l(l+1)}{2} \left( 1- \frac{1}{b^2} \right) \right\rbrace \psi
+ b\dot{b}\dot{\chi} + \left( b\dot{b} - \frac{\dot{b}}{b} \right) \dot{\psi} + \ddot{\psi} =0,
\end{equation}

\begin{equation}\label{25f}
\begin{split}
(\dot{a}^2b^4 + a^2b^3\ddot{b})(\chi+\psi) + (a^4\dot{b}^2 + a^3\ddot{a}b^2)\psi
+ (-a\dot{a}b^4 + \frac{1}{2}a^4b^3\dot{b} + \frac{1}{2}a^3\dot{a}b^4 + \frac{1}{2}a^2b^3\dot{b}) (\dot{\chi}+\dot{\psi}) + ( \frac{1}{2}a^3\dot{a}b^2 - a^4b\dot{b}) \dot{\psi} \\
 + \frac{1}{2} a^2b^4 (\ddot{\chi} +\ddot{\psi} +\chi'' +\psi'')
+ \frac{1}{2}a^4b^2 \ddot{\psi} - \frac{1}{2}a^2b^2 \psi''  
 - a^2b^3\dot{b}\zeta' - a^2b^4\dot{\zeta}' + \frac{1}{2Y}a^2b^2(a^2 -1)(\chi+\psi) \frac{1}{\theta} (\partial_{\theta}Y) =0,
\end{split}
\end{equation}

\begin{equation}\label{25g}
\begin{split}
(\dot{a}^2b^4 + a^2b^3\ddot{b})(\chi+\psi) + (a^4\dot{b}^2 + a^3\ddot{a}b^2)\psi
+ (-a\dot{a}b^4 + \frac{1}{2}a^4b^3\dot{b} + \frac{1}{2}a^3\dot{a}b^4 + \frac{1}{2}a^2b^3\dot{b}) (\dot{\chi}+\dot{\psi}) + ( \frac{1}{2}a^3\dot{a}b^2 - a^4b\dot{b})\dot{\psi} \\
 + \frac{1}{2} a^2b^4 (\ddot{\chi} +\ddot{\psi} +\chi'' +\psi'')
+ \frac{1}{2}a^4b^2 \ddot{\psi} - \frac{1}{2}a^2b^2 \psi''  
- a^2b^3\dot{b}\zeta' - a^2b^4\dot{\zeta}' + \frac{1}{2Y}a^2b^2(a^2 -1)(\chi+\psi) (\partial_{\theta} \partial_{\theta}Y) =0.
\end{split}
\end{equation}

Adding equations \eqref{25f} and \eqref{25g}, and simplifying gives
\begin{equation}\label{25h}
\begin{split}
\left\lbrace b\ddot{b} + \frac{\dot{a}^2 b^2}{a^2} -\frac{l(l+1)}{4}(a^2 -1) \right\rbrace \chi
+ \left\lbrace a\ddot{a} + b\ddot{b} + \frac{\dot{a}^2 b^2}{a^2} + \frac{a^2 \dot{b}^2}{b^2} -\frac{l(l+1)}{4}(a^2 -1) \right\rbrace \psi \\
+ \left\lbrace - \frac{\dot{a} b^2}{a} +\frac{1}{2} b\dot{b} +\frac{1}{2} a\dot{a} b^2 +\frac{1}{2} a^2 b\dot{b} \right\rbrace \dot{\chi}
+ \left\lbrace - \frac{\dot{a} b^2}{a} - \frac{a^2 \dot{b}}{b} + \frac{1}{2} a\dot{a} +\frac{1}{2} b\dot{b} +\frac{1}{2} a\dot{a} b^2 +\frac{1}{2} a^2 b\dot{b} \right\rbrace  \dot\psi \\
+ \frac{1}{2} b^2(\ddot{\chi}+\chi'') + \frac{1}{2} (a^2+b^2) \ddot{\psi} + \frac{1}{2} (b^2 -1) \psi''
- b\dot{b}\zeta' - b^2\dot{\zeta}' =0.
\end{split}
\end{equation}

\subsection{In vacuum: Solutions to Perturbation equations}

In order to eliminate the $ \dot{\zeta}' $-term from Eqn.\eqref{25h},
Eqn.\eqref{25c} is differentiated w.r.t. $ r $, which yields
\begin{equation}\label{25i}
\dot{\zeta}'= -\frac{\dot{a}}{a}\zeta' -\chi'' +\left(\frac{1}{b^2}-1\right) \psi'',
\end{equation}
and Eqn.\eqref{25h} now reads as
\begin{equation}\label{25j}
\begin{split}
\left\lbrace b\ddot{b} + \frac{\dot{a}^2 b^2}{a^2} -\frac{l(l+1)}{4}(a^2 -1) \right\rbrace \chi
+ \left\lbrace a\ddot{a} + b\ddot{b} + \frac{\dot{a}^2 b^2}{a^2} + \frac{a^2 \dot{b}^2}{b^2} - \frac{l(l+1)}{4}(a^2 -1) \right\rbrace \psi \\
+ \left\lbrace - \frac{\dot{a} b^2}{a} +\frac{1}{2} b\dot{b} +\frac{1}{2} a\dot{a} b^2 +\frac{1}{2} a^2 b\dot{b} \right\rbrace \dot{\chi}
+ \left\lbrace - \frac{\dot{a} b^2}{a} - \frac{a^2 \dot{b}}{b} + \frac{1}{2} a\dot{a} +\frac{1}{2} b\dot{b} +\frac{1}{2} a\dot{a} b^2 +\frac{1}{2} a^2 b\dot{b} \right\rbrace  \dot\psi \\
+ \frac{1}{2} b^2 \ddot{\chi} - \frac{1}{2} b^2 \chi'' + \frac{1}{2} (a^2+b^2) \ddot{\psi} - \frac{1}{2} (b^2 -1) \psi'' + \left\lbrace \frac{\dot{a} b^2}{a} - b\dot{b} \right\rbrace \zeta' =0.
\end{split}
\end{equation}

This equation \eqref{25j}, having both the $ \ddot{()}$ and $ ()''$ terms, resembles a wave equation in $ \chi (t,r) $ and $ \psi (t,r) $. One can see that unlike the axial case, the solutions for polar modes cannot be derived easily even in the vacuum case for $ l \geq 2 $. However, we may proceed further only if, after inserting $ l=2 $ and substituting Eqn.\eqref{6}, we make certain assumptions, such as neglecting the perturbation due to $ \chi (t,r) $ (or $ \psi (t,r) $), or assuming $ \zeta (t,r) $ to be constant, or assuming $ \chi (t,r) $ and $ \psi (t,r) $) to be proportional to each other. Before proceeding with these assumptions, we write down the polar perturbation equations which will be used in the subsequent analysis:
\begin{equation}\label{26a}
- 3 \zeta + \frac{2}{3} K ^2 t^{1/3} \chi' + \left(\frac{2}{3} K ^2 t^{1/3} - \frac{1}{3t} \right) \psi' + \dot{\psi}' =0,
\end{equation}
\begin{equation}\label{26c}
- \frac{1}{3t} \zeta + \dot{\zeta} - \chi' + \left(\frac{1}{K^2 t^{4/3}} -1\right) \psi' =0,
\end{equation}
and
\begin{equation}\label{26e}
-3 \chi  + \left( \frac{4}{9t^2} + \frac{3}{K^2 t^{4/3}} -3 \right) \psi
+ \frac{2}{3} K^2 t^{1/3} \dot{\chi} + \left( \frac{2}{3} K^2 t^{1/3} -\frac{2}{3t} \right) \dot{\psi} + \ddot{\psi} =0.
\end{equation}
We now attempt to analyse these equations in particular cases.
\bigskip

\underline{Case 1}: \\
Let us assume that $ \chi (t,r) $ and $ \psi (t,r) $ are proportional to each other, i.e.
\begin{equation}\label{27a}
\psi (t,r) =q \chi (t,r),
\end{equation}
$ q $ being a constant, then Eqn.\eqref{26e} leads to:
\begin{equation}\label{27b}
\left( -3 - 3q + \frac{4q}{9t^2} + \frac{3q}{K^2 t^{4/3}} \right) \chi + \left( \frac{2}{3} K^2 t^{1/3} + \frac{2}{3} qK^2 t^{1/3} - \frac{2q}{3t} \right) \dot{\chi} + q \ddot{\chi} = 0.
\end{equation}
Its solution is given by
\begin{equation}\label{27c}
\begin{split}
\chi (t,r) 
= t^{1/3}\exp \left( \dfrac{27t^{2/3}}{4K^2} \right) (t F_1(r) B_1 + F_2(r) B_2) ,
\end{split}
\end{equation}
where
\begin{equation*}
\begin{split}
B_1 = \text{HeunB} \hspace{0.1cm} \left( \frac{3}{2}, \hspace{0.1cm} \frac{54q^2 (1+q)}{(-2q (1+q))^{3/2} K^3}, \hspace{0.1cm}
- \frac{4(1+q) K^6 - 729q}{8K^6(1+q)}, \hspace{0.1cm} -\frac{27q}{K^3 \sqrt{-2q (1+q)}},
 \hspace{0.1cm} \frac{\sqrt{-2q (1+q)} Kt^{2/3}}{2q} \right) ,
\end{split}
\end{equation*}
\begin{equation}\label{27d}
\begin{split}
B_2 = \text{HeunB} \hspace{0.1cm} \left( -\frac{3}{2}, \hspace{0.1cm} \frac{54q^2 (1+q)}{(-2q (1+q))^{3/2} K^3}, \hspace{0.1cm}
- \frac{4(1+q) K^6 - 729q}{8K^6(1+q)}, \hspace{0.1cm} -\frac{27q}{K^3 \sqrt{-2q (1+q)}},
 \hspace{0.1cm} \frac{ \sqrt{-2q (1+q)} Kt^{2/3} }{2q} \right).
\end{split}
\end{equation}
$ B_1 $ and $ B_2 $ are two different biconfluent Heun's functions, and $ F_1(r) $ and $ F_2(r) $ are undetermined functions of $ r $.
Hence, Eqn.\eqref{27a} gives
\begin{equation}\label{27e}
\begin{split}
\psi(t,r) 
= q t^{1/3}\exp \left( \dfrac{27t^{2/3}}{4K^2} \right) (t F_1(r) B_1 + F_2(r) B_2) .
\end{split}
\end{equation}

\begin{figure}[h]
\centering
\begin{minipage}[b]{0.47\linewidth}
\includegraphics[scale=0.33]{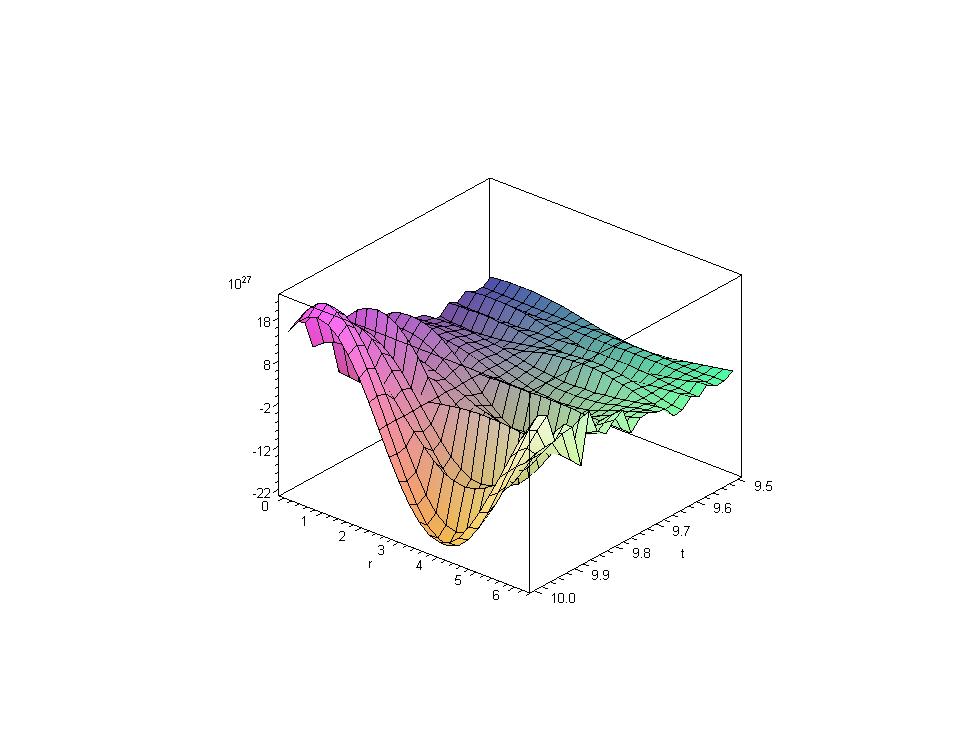}
\begin{center} (a)
\end{center}
\end{minipage}
\begin{minipage}[b]{0.47\linewidth}
\includegraphics[scale=0.33]{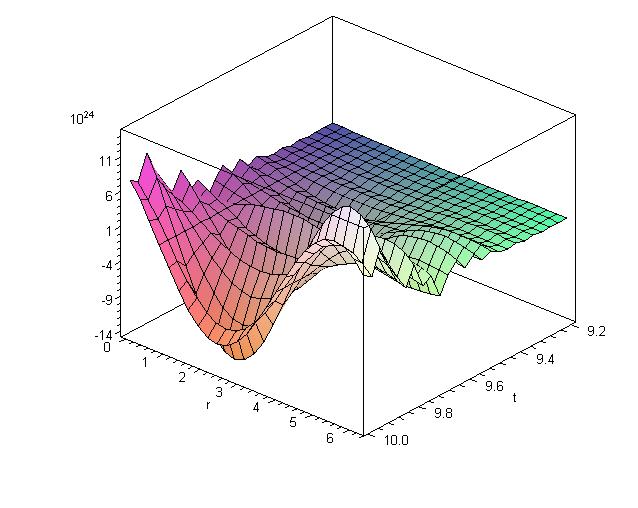}
\begin{center} (b)
\end{center}
\end{minipage}
\caption{The vertical axes represent the polar perturbations in Case 1: (a) $ \chi (t,r) $ given by Eqn.\eqref{27c}, and (b) $ \zeta (t,r) $ given by Eqn.\eqref{27f}. } \label{fig:Plot5} 
\end{figure}

Moreover, from Eqn.\eqref{26a}, we get
\begin{equation}\label{27f}
\begin{split}
\zeta (t,r) = \frac{1}{9K^2} \exp \left( \dfrac{27t^{2/3}}{4K^2} \right)
\left[ \frac{2}{t^{5/3}} \left\lbrace (1+q) K^4 t^{10/3} + \frac{27}{4} q t^{8/3} + \frac{3}{2} qK^2 t^2 \right\rbrace   B_1 F'_1
\right. \\ \left.
+ 2 \left\lbrace  (1+q) K^4 t^{2/3} + \frac{27}{4} q \right\rbrace B_2 F'_2
+ K^3 \sqrt{-2q (1+q)} \left\lbrace t \tilde{B}_1 F'_1
+ \tilde{B}_2 F'_2  \right\rbrace \right].
\end{split}
\end{equation}
The tilde over B's denotes the $ z $-derivative of the biconfluent Heun's function, $\textrm{HeunB}(\alpha , \beta , \gamma , \delta , z )$.
The variations of $ \chi (t,r) $ in Eqn.\eqref{27c}, and $ \zeta (t,r) $ in Eqn.\eqref{27f} are shown in Fig.\eqref{fig:Plot5}.
\\

\underline{Case 2}: \\
Putting $ \chi (t,r) =0 $, Eqn.\eqref{26e} reduces to:
\begin{equation}\label{28a}
\left( -3 + \frac{4}{9t^2} + \frac{3}{K^2 t^{4/3}} \right) \psi + \left( \frac{2}{3} K^2 t^{1/3} - \frac{2}{3t} \right) \dot{\psi} + \ddot{\psi} = 0,
\end{equation}
which is solved by
\begin{equation}\label{28b}
\begin{split}
\psi (t,r) 
 = t^{1/3} \exp \left( -\frac{t^{2/3} (K^4 t^{2/3} +27) }{4K^2} \right) (t F_3(r) B_3 + F_4(r) B_4),
\end{split}
\end{equation}
with
\begin{eqnarray}\label{28c}
B_3 = \text{HeunB} \hspace{0.1cm} \left( \frac{3}{2}, \hspace{0.1cm} \frac{27 \sqrt{2}}{2K^3}, \hspace{0.1cm} \frac{(4K^6 +729)}{8K^6},
\hspace{0.1cm} -\frac{27 \sqrt{2}}{2K^3}, \hspace{0.1cm} \frac{ \sqrt{2} Kt^{2/3} }{2} \right),   \\
B_4 = \text{HeunB} \hspace{0.1cm} \left( -\frac{3}{2}, \hspace{0.1cm} \frac{27 \sqrt{2}}{2K^3}, \hspace{0.1cm} \frac{(4K^6 +729)}{8K^6},
\hspace{0.1cm} -\frac{27 \sqrt{2}}{2K^3}, \hspace{0.1cm} \frac{ \sqrt{2} Kt^{2/3} }{2} \right).
\end{eqnarray}

Inserting this in Eqn.\eqref{26a} gives 
\begin{equation}\label{28d}
\begin{split}
\zeta (t,r) = \frac{1}{18K^2 t^{5/3}} \exp \left( -\frac{t^{2/3} (K^4 t^{2/3} +27) }{4K^2} \right)  \left[  \left\lbrace 6K^2 t^2 - 27 t^{8/3} \right\rbrace B_3 F'_3
-27 t^{5/3} B_4 F'_4
\right. \\ \left.
+ 2\sqrt{2}K^3 \left\lbrace t^{8/3}  \tilde{B}_3 F'_3 + t^{5/3} \tilde{B}_4 F'_4 \right\rbrace   \right].
\end{split}
\end{equation}
Here $ F_3(r) $ and $ F_4(r) $ are undetermined functions of $ r $. As in the previous case, two different biconfluent Heun's functions represented by $ B_3 $ and $ B_4 $ appear in the $ t $-solution. The tilde over `$ B $'- s  indicates the respective $ z $-derivatives.  \\

\begin{figure}[h]
\centering
\begin{minipage}[b]{0.45\linewidth}
\includegraphics[scale=0.3]{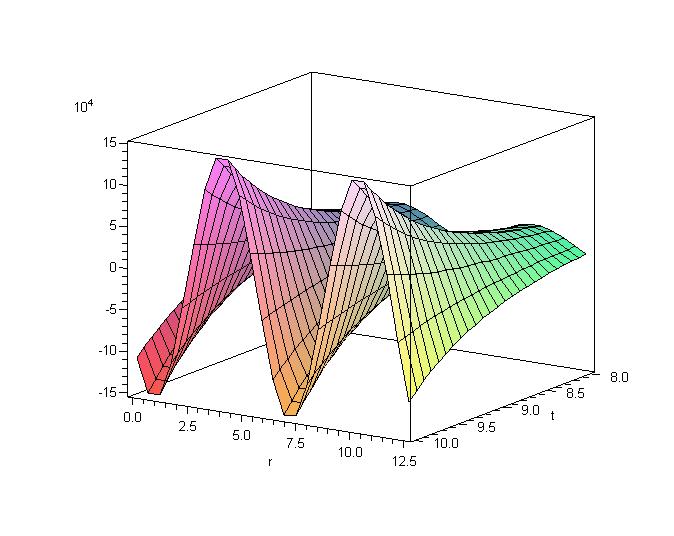}
\begin{center} (a)
\end{center}
\end{minipage}
\begin{minipage}[b]{0.45\linewidth}
\includegraphics[scale=0.3]{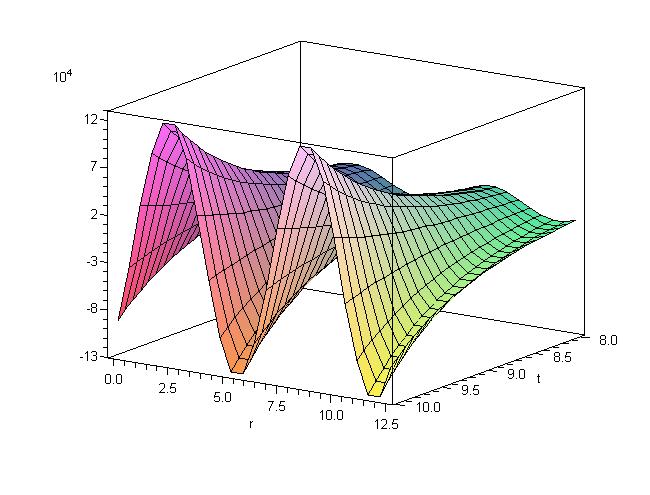}
\begin{center} (b)
\end{center}
\end{minipage}
\caption{The vertical axes represent (a) $ \psi (t,r) $ from Eqn.\eqref{28b}, and (b) $ \zeta (t,r) $ from Eqn.\eqref{28d} in Case 2 of polar solutions. } \label{fig:Plot6} 
\end{figure}


\bigskip

\underline{Case 3}: \\
Lastly, we choose $ \zeta (t,r) $= constant =1. This choice eliminates the derivatives of $ \zeta $ from the perturbation equations. We will show in \textit{Discussions II} that Eqn.\eqref{25j} under this particular assumption reduces to an equivalent of the Zerilli equation. From Eqn.\eqref{26c}, we get
\begin{equation}\label{32a}
\chi'= - \frac{1}{3t} + \left(\frac{1}{K^2 t^{4/3}} -1\right) \psi'.
\end{equation}
Integrating w.r.t. $ r $ and setting the integration constant to zero, Eqn.\eqref{32a} gives
\begin{equation}\label{32b}
\chi (t,r)= - \frac{r}{3t} + \left(\frac{1}{K^2 t^{4/3}} -1\right) \psi (t,r),
\end{equation}
which when substituted in Eqn.\eqref{26e} yields an equation in $ \psi (t,r) $ only:
\begin{equation}\label{32d}
\ddot{\psi} - \dfrac{4}{9t^2} \psi + \dfrac{2}{9} \dfrac{K^2 r}{t^{5/3}} + \frac{r}{t} =0.
\end{equation}
Hence, we arrive at
\begin{equation}\label{32e}
\psi (t,r) = F_6(r) t^{4/3} + F_7(r) t^{-1/3} + \frac{1}{12} r t^{1/3} (27t^{2/3} + 4K^2),
\end{equation}      
\begin{equation}\label{32f}
\chi (t,r)= - \frac{r}{3t} + \left(\frac{1}{K^2 t^{4/3}} -1\right) \left[ F_6(r) t^{4/3} + F_7(r) t^{-1/3} + \frac{1}{12} r t^{1/3} (27t^{2/3} + 4K^2) \right].
\end{equation}
Here also, we obtain $ F_6(r) $ and $ F_7(r) $ as undetermined functions of $ r $.

\begin{figure}[h]
\centering
\begin{minipage}[b]{0.47\linewidth}
\includegraphics[scale=0.32]{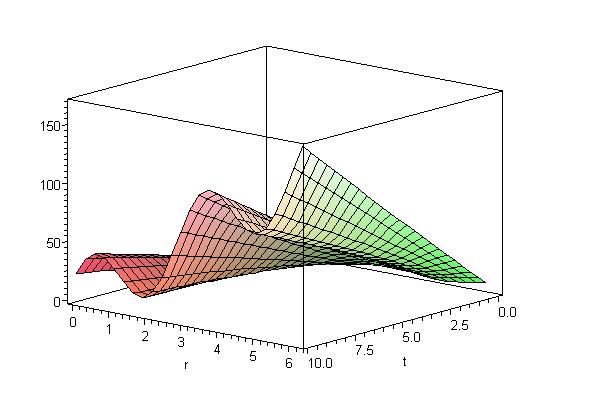}
\begin{center} (a)
\end{center}
\end{minipage}
\begin{minipage}[b]{0.47\linewidth}
\includegraphics[scale=0.33]{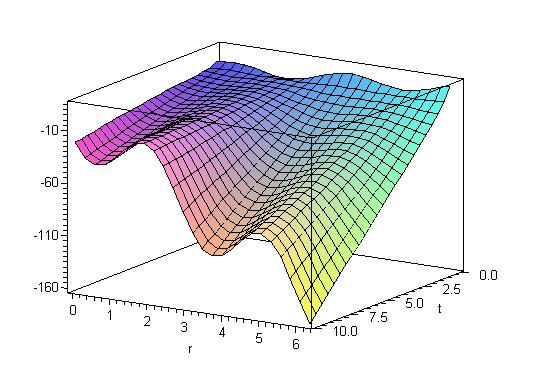}
\begin{center} (b)
\end{center}
\end{minipage}
\caption{The vertical axes represent the polar perturbations in Case 3: (a) $ \psi (t,r) $ given by Eqn.\eqref{32e}, and (b) $ \chi (t,r) $ given by Eqn.\eqref{32f}. } \label{fig:Plot8} 
\end{figure}

\subsubsection*{Discussions II}

In all the cases discussed here, we find that the radial part of the solutions remain undetermined. However, on inspection, it can be shown from the set of perturbation equations that 
\begin{equation}\label{30a}
\zeta (t,r) \propto \mathcal{R}_{\mu}'(r) \hspace{0.6cm} \text{or} \hspace{0.6cm} \zeta (t,r) \propto \int \mathcal{R}_{\mu}(r) dr , \hspace{1cm} \mu = \chi, \psi .
\end{equation}
$ \mathcal{R}(r) $ is the radial part of the perturbations $ \chi (t,r) $ and $ \psi (t,r) $. This indicates that $ \mathcal{R}(r) $ will behave sinusoidally. Hence, one can express it in the form:
\begin{equation}\label{30b}
\mathcal{R}_{\mu} (r) = \lambda_1 \sin (Mr) + \lambda_2 \cos (Mr).
\end{equation}
Subsequently,
\begin{equation}\label{30c}
| \mathcal{R}_{\zeta} (r)| = \lambda_3 \left[\lambda_1 \cos (Mr) - \lambda_2 \sin (Mr)\right].
\end{equation}
Here, $ \lambda_1 $, $ \lambda_2 $, $ \lambda_3 $ and $ M $ are constants. Since the $ t $ and $ r $-solutions separate as a product in the Regge-Wheeler formalism, the polar perturbation terms can be split as:
\begin{equation}\label{30d}
\nu (t,r) = \mathcal{T}_{\nu} (t) \mathcal{R}_{\nu} (r), \hspace{1cm} \nu =\chi, \psi, \zeta,
\end{equation}
the $ \mathcal{T}_{\nu} (t) $ solutions being explicitly obtained in equations \eqref{27c}-\eqref{27f}, \eqref{28b}-\eqref{28d}, 
 \eqref{32e}-\eqref{32f}, and the $ \mathcal{R}_{\nu} (r) $ solutions in \eqref{30b} and \eqref{30c}.
However, in Case 3, where $ \zeta (t,r) $= constant =1, the solution has a slightly different nature. In addition to the terms containing undetermined functions of $ r $, another term involving only $ r $ multiplied by some powers of $ t $ appears here. The temporal part of the solution is much simpler as compared to that in the other cases. Inserting suitable values for the constants, we have generated the 3-dimensional plots (Figs.\eqref{fig:Plot5}-\eqref{fig:Plot8}) of the perturbations for each set of the polar solutions.




We can determine an order-of-magnitude estimate of the frequency of the polar waves to be lying approximately in the range 1000-2000 Hz from the plot of the temporal part of the $ \chi (t,r) $-solution in Case 1 with $ q = K = 1 $. The strain generated by the GWs of a given frequency can be used to constrain the perturbation parameters in the same way as we did for Kantowski-Sachs background (Sec.7 of \cite{GD2}). We find that the strain in Bianchi I spacetime (with $ \theta =\pi /45 $) is roughly four times the strain obtained in the Kantowski-Sachs spacetime \cite{GD2}. The constraints on the unknown constants are found to be similar in both the spacetimes, so as to keep the magnitude of strain to be in agreement with the observational data: $ \sim 10^{-24}$ \cite{PAPA1,PAPA2} for the specified frequency range.

Now, from Eqn.\eqref{25j}, we have 
\begin{equation}\label{26f}
\begin{split}
\left( \dfrac{3}{2} - \frac{3}{2K t^{2/3}} - \frac{K^2}{9 t^{2/3}} \right) \chi
+ \left( \dfrac{3}{2} - \frac{3}{2K t^{2/3}} + \frac{8}{9K t^{8/3}} - \frac{K^2}{9 t^{2/3}} \right) \psi
+ \left( \frac{2}{3} K^2 t^{1/3} + \frac{K}{6t^{1/3}} \right) \dot{\chi} \\
+ \left( \frac{2}{3} K^2 t^{1/3} + \frac{K}{6t^{1/3}} - \frac{5}{6K t^{5/3}} \right) \dot\psi
 - K^2 t^{1/3} \zeta'
+ \frac{1}{2} K^2 t^{4/3} \ddot{\chi} + \frac{1}{2} \left( K^2 t^{4/3} + \frac{1}{K t^{2/3}} \right) \ddot{\psi} \\
- \frac{1}{2} K^2 t^{4/3} \chi'' - \frac{1}{2} (K^2 t^{4/3} -1) \psi'' =0.
\end{split}
\end{equation}
Using the relation \eqref{32b} between $ \chi (t,r) $ and $ \psi (t,r) $ when $ \zeta (t,r) $ is assumed to be unity, Eqn.\eqref{26f} reads as
\begin{equation}\label{32c}
\dfrac{1}{2}\left( 1+ \dfrac{1}{K t^{2/3}} \right) \ddot{\psi} - \dfrac{2}{3} \left( \dfrac{1}{t} + \dfrac{1}{K t^{5/3}} \right) \dot{\psi} + \left( \dfrac{3}{2K^2 t^{4/3}} + \dfrac{2}{3K t^{8/3}} - \dfrac{3}{2K^3 t^2} + \dfrac{5}{9t^2} \right) \psi + \left \lbrace \dfrac{Kr}{18 t^{7/3}} + \dfrac{r}{2K t^{5/3}} - \dfrac{2K^2 r}{27t^{5/3}} - \dfrac{r}{2t} \right \rbrace =0.
\end{equation}

This equation involving only one polar perturbation variable can be said to be the analogue of the Zerilli equation for the Bianchi I background. The pre-factor of $ \psi (t,r) $ behaves as a potential. One can compare it with Eq.(101) of Ref.\cite{REZ}. An additional $ \dot{\psi} $-term is responsible for damping. However, the terms containing the second-order $ r $-derivatives of $ \psi (t,r) $ disappear here. The equation is inhomogeneous due to the presence of the terms within the curly brackets.

\section{Conclusions}

We have investigated the axial and polar perturbations to a matter-filled Bianchi I spacetime using the Regge-Wheeler gauge. To begin with, we derived the solutions of the background field equations. The relation $ a= b^n $ among the scale factors imply a matter content which behaves like a stiff perfect fluid. We have solved for the scale factors, the fluid pressure and energy density as explicit functions of time. Furthermore, the numerical value of $ n $ can be exactly determined for the stiff fluid as well as for the vacuum case from the conditions proposed by Jacobs \cite{WAINBOOK}. The two different values of $ n $ subsequently leads to somewhat different nature of the GW solutions in the presence and absence of matter in Bianchi I spacetime. No such condition similar to \cite{WAINBOOK} was available for determining the Kantowski-Sachs background solutions. The $ n $-value remained arbitrary for the stiff fluid (we assumed $ n= 1/2 $), but was obtainable for the vacuum background \cite{GD2}.

Moving on to the perturbed metric, the axial and polar perturbation equations are treated separately. In either case, these equations yield the $ t $ and $ r $-solutions of the perturbing terms. The $ \theta $-dependence is defined by the term $ \sin \theta (\partial_{\theta}Y) $ in the axial case \eqref{7a} and by $ Y (\theta) $ in the polar case \eqref{7e}.
In accordance with \cite{RW}, the $ \phi $-dependence has been removed at the beginning by choosing $ m=0 $. Thus every perturbing element can be expressed by a product of four terms, each being a function of only one of the coordinates $ t $, $ r $, $ \theta $ and $ \phi $.

In Sec.V, we have dealt with the axial perturbations. First we derived the linearised field equations for the axially perturbed metric \eqref{7b}. Then we obtained the wave equation and the respective (semi-)analytical solutions in terms of $ \mathcal{Q} (t,r) $, from which the axial perturbations $ h_0 (t,r) $ and $ h_1 (t,r) $ are evaluated. As pointed out earlier, Eqn.\eqref{13d} can be called the Regge-Wheeler equation in Bianchi I background. In the wave equation \eqref{13d}, the pre-factor of $ \mathcal{Q} $ serves as an effective potential (analogous to Ref.\cite{RW}, Ref.\cite{VISH}, Ref.\cite{REZ}). Therefore, it is nothing but a wave equation in a scattering potential, as mentioned by Rezzolla \cite{REZ}. Both the first-order and second-order derivatives of $ \mathcal{Q} $ are present in these equations. The terms $ \dot{\mathcal{Q}} $ and $ \mathcal{Q}' $ are responsible for the damping of the axial GWs travelling in the $ (t-r) $ hypersurface. We can infer that this damping originates from the anisotropy of the Bianchi I background. It is known that unlike in FLRW universe \cite{M1, SH1}, damping terms appear in LTB background \cite{CL1}. Such damping has been found to occur in Kantowski-Sachs spacetime \cite{GD2} also.

In all cases, the solutions of $ \mathcal{Q} (t,r) $, and hence $ h_0 (t,r) $ and $ h_1 (t,r) $ in the matter-filled Bianchi I spacetime are very similar. They are in the form of combinations of hypergeometric and modified Bessel functions of first kind. To arrive at physical solutions, one has to consider the real part of the imaginary terms occurring here. When $ \alpha $ (the power of $ r $), is non-zero, irrespective of the azimuthal fluid velocity $ U $,
$ r $ appears as a multiplicative factor in the denominator. For different $ U $-values, but the same set of $ \alpha $ and $ \beta $,
the difference is due to the presence of $ r $ as a coefficient of the second term (sum of hypergeometric and Bessel functions) in the numerator.
Making assumptions for determining $ U $, in particular the normalisation condition \eqref{N}, is a crucial step in finding complete solutions for the perturbations $ h_0(t,r) $ and $ h_1(t,r) $. Its value depends explicitly on the product of $ b(t) $, and some functions $ T(t) $ and $ R(r) $.
Sharif and his collaborators derived the expression for the azimuthal velocity from the solutions of the axial perturbations \cite{SH1}. However, in Bianchi I spacetime, the equations are not as simple as in \cite{SH1} because of the different scale factors, and we need to assume possible expressions for $ U $ in order to determine the axial solutions.
The solutions become easier to derive when matter is absent. The expressions for $ h_0(t,r) $ and $ h_1(t,r) $ are obtained in terms of Heun's biconfluent functions (temporal part) and sinusoidal functions (radial part) (see Ref.\cite{GD1}). Unlike in the case of matter-filled spacetime, the vacuum perturbation equations do not require additional assumptions and can be solved by the method of separation of variables.

We have used $ l=2 $ for wave-like solutions in both the axial and polar cases. For $ l = 0 $, the spherical harmonics are characterised by spherical symmetry. However, $ l \geq 0 $ indicates deviation from spherical symmetry, which renders the quadrupole moment non-zero, and hence important in gravitational radiations. According to Clarkson \textit{et al.} \cite{CL1}, scalars on $ S^2 $ can be expressed as a sum over polar modes, and higher-rank tensors as sums over both the polar and axial modes. We need to consider $ l \geq 2 $ in order to take into account the axial modes coming from the expansion of both vector and tensor functions. The value of $ l $ determines the height of the effective potential barrier, given by the coefficient of $ \mathcal{Q} $ in the wave equation for axial modes \cite{RW}. A change in the value of $ l $ brings about only a small change in the Heun's function, without affecting the radial solution \cite{GD1}. Consequently, the expressions for $ h_0(t,r) $ and $ h_1(t,r) $ will slightly change.
Although no axial perturbations with $ l = 0 $ are feasible \cite{CL1}, there exist polar perturbations for all values: $ l=0,1,2 \cdots $. In the $ (0-0) $ element of the polar perturbation matrix (Eqn.\eqref{7e}), a third term $ \eta (t,r) $ appears in addition to $ \chi (t,r) $ and $ \psi (t,r) $. Studies have shown that $ \eta $ vanishes in the FLRW background \cite{M2, ROST2}. But Clarkson \textit{et al.} \cite{CL1} remarked that when $ l=0 $ or 1, i.e. for large angle fluctuations, the field equations do not yield $ \eta =0 $. The assumption $ \eta =0 $ itself acts as a constraint equation \cite{CL1}.  We have treated the zero and non-zero values of $ \eta $ in separate subsections.

The stiff fluid conditions have been used to simplify the polar perturbation equations in presence of matter. In this case, under certain assumptions, we have been able to extract a solution (Eqn.\eqref{24b}) for the polar perturbation $ \chi (t,r) $. However, this solution carries an undetermined function depending on both $ t $ and $ r $, i.e., the temporal and radial parts do not separate out as products and therefore cannot be considered as a valid solution in the context of RW gauge.
Since the absence of matter further simplifies the perturbation equations, we have analysed the vacuum case in details. We find that the polar perturbation equations, even with $ \eta =0 $, contain far more complicated couplings among the perturbing elements than the axial perturbations to the Bianchi I background (and Kantowski-Sachs backrgound) and also their FLRW counterparts. Unlike in the case of FLRW background \cite{M2, SH2, ROST2}, no perturbation equation in a single variable can be obtained in the Bianchi I case, thereby complicating the derivation of analytical solutions for the polar modes. However, if we assume a constant value for the perturbation variable $ \zeta (t,r) $, we can extract an equation in only $ \psi (t,r) $ (please see Eqn.\eqref{32c}), which is comparable to the Zerilli equation \cite{REZ} for polar perturbations.
Following Clarkson \textit{et al.} \cite{CL1}, we have solved the equations analytically in some particular cases, such as assuming $ \chi (t,r) $ and $ \psi (t,r) $ proportional to each other, or $ \zeta (t,r) $ = constant, or switching off either $ \chi (t,r) $ or $ \psi (t,r) $.  In each of these cases, the polar solutions may be expressed as a product of a radial function which is sinusoidal in nature, and a temporal function which is again a combination of biconfluent Heun's functions and their derivatives (Eqns.\eqref{27c}-\eqref{27f}, \eqref{28b}-\eqref{28d}),
or some powers of $ t $. In the last case, the polar solution is much simpler compared to the other two. The figures in \eqref{fig:Plot5}-\eqref{fig:Plot8} show the nature of these solutions. The combined effects of these complicated perturbing terms will result in the spacetime perturbations.

The effects of axial and polar GWs on the material medium, which are already established in the FLRW background \cite{M2, SH1, SH2, SIDD}, have been found to hold in the Bianchi I background also. The azimuthal velocity of the fluid is perturbed by the axial modes only. The non-zero value of azimuthal velocity $ U $ indicates the cosmological rotation of the fluid induced by the propagating axial GWs. The remaining components are influenced by the polar GWs. Besides, the energy density and pressure undergo deformation due to the polar modes but remain unaffected in the presence of axial waves. We have shown that polar modes cannot exist if the matter perturbations vanish. Thus the polar GWs must be followed up by matter inhomogeneities and anisotropies.
The same effects have been demonstrated in our studies on Kantowski-Sachs background \cite{GD2}. Comparing the corresponding results in Bianchi I and Kantowski-Sachs spacetimes, we find that both the axial and polar GWs are much alike in nature. The nature of temporal and radial solutions of the axial perturbation equations in vacuum are very similar. Due to the difference in the term containing $ l $ in the wave equation, a minor difference appears only in the Heun's function. The polar solutions in vacuum also closely resemble one another, except for differences in the numerical factors within the exponential and Heun's functions. The assumption that $ \cot \theta \simeq 1/ \theta $, valid for small values of $ \theta $, has been made in Bianchi I case, but is not required in the KS background. This is because of the nature of the line element of the background spacetime.

It has been shown through rigorous calculations in \cite{CL1} that the gauge-invariant perturbation parameters in the inhomogeneous LTB model comprise of a cumbersome mixture of scalar, vector and tensor modes. Their couplings further complicate the evolution equations. Thus these gauge-invariants are much different from those in FLRW model. On the other hand, spherically symmetric spacetimes do not favour SVT decomposition \cite{MEYER}. Likewise, we observe that the comparison of the corresponding results in Bianchi I and FLRW spacetimes is non-trivial. Equating the two scale factors responsible for anisotropy is not sufficient to reproduce the results in the isotropic limit because we cannot extract the master equation in terms of a single variable. Only when considered in a general gauge chosen suitably as in \cite{CL1}, can the perturbations in the two models be matched. We are presently working on the general gauge so as to match the two perturbations.


\section*{Acknowledgement}
SD acknowledges the financial support from INSPIRE (AORC), DST, Govt. of India (IF180008). SG thanks IUCAA, India for an associateship and CSIR, Government of India for the major research grant [No. 03(1446)/18/EMR-II]. SC is grateful to CSIR, Government of India for providing fellowship.


\begin{thebibliography}{}
\bibitem{calc} Some calculations in this paper are done using Maple software, and the rest by hand.
\end{thebibliography}
\end{document}